\documentclass[a4paper,twocolumn,11pt,superscriptaddress,nofootinbib,accepted=2025-03-28]{quantumarticle}
\pdfoutput=1
\usepackage{float}
\makeatletter
\let\newfloat\newfloat@ltx
\makeatother
\usepackage[utf8]{inputenc}
\usepackage[english]{babel}
\usepackage[T1]{fontenc}
\usepackage{amsmath}
\usepackage{amsfonts}
\usepackage{bm}
\usepackage{mathrsfs}
\usepackage{latexsym}
\usepackage{graphicx}

\usepackage{braket}
\usepackage{amsthm}
\usepackage{hyperref}
\usepackage{enumitem}
\usepackage{multirow}
\usepackage{makecell}
\usepackage{comment}
\usepackage{booktabs}
\usepackage{siunitx}

\allowdisplaybreaks

\hypersetup{colorlinks=true,}

\newcommand*\chem[1]{\ensuremath{\mathrm{#1}}}

\usepackage{algpseudocode}
\usepackage{algorithm}
\usepackage{footnotebackref}
\algrenewcommand\alglinenumber[1]{\sf\scriptsize\color{blue}{#1}}
\algrenewcommand\algorithmicrequire{\textbf{Input:}}
\algrenewcommand\algorithmicensure{\textbf{Output:}}

\makeatletter
\newcommand{\Oplus}{\mathbin{\mathpalette\make@big\oplus}}

\newcommand{\make@big}[2]{%
  \vcenter{\hbox{%
    \scalebox{1.3}{$\m@th#1#2$}%
  }}%
}
\makeatother
\usepackage{cleveref}
\DeclareSymbolFont{extraup}{U}{jkpsyc}{m}{n}
\DeclareMathSymbol{\vardiamondsuits}{\mathalpha}{extraup}{113}

\newtheorem{theorem}{Theorem}
\newtheorem{facts}{facts}
\theoremstyle{remark}

\begin{document}
\title{Resource-Optimized Grouping Shadow for Efficient Energy Estimation}
\author{Min Li}
\affiliation{Department of Physics, University of Illinois Urbana-Champaign, Urbana, IL 61801, USA}
\author{Mao Lin}
\author{Matthew J.~S.~Beach}
\affiliation{AWS Quantum Technologies, Seattle, WA 98170, USA}
\begin{abstract}
\onecolumn

\begin{minipage}{0.95\textwidth}

The accurate and efficient energy estimation of quantum Hamiltonians consisting of Pauli observables is an essential task in modern quantum computing. We introduce a Resource-Optimized Grouping Shadow (ROGS) algorithm, which optimally allocates measurement resources by minimizing the estimation error bound through a novel overlapped grouping strategy and convex optimization. Our numerical experiments demonstrate that ROGS requires significantly fewer unique quantum circuits for accurate estimation accuracy compared to existing methods given a fixed measurement budget, addressing a major cost factor for compiling and executing circuits on quantum computers. 
\end{minipage}
\end{abstract}

%%%%%%%%%%%%%%%%%%%%%%%%%%%%%%%%%%%%

\maketitle

\nopagebreak

\section{Introduction}
\twocolumn
Recent advancements in Noisy Intermediate-Scale Quantum (NISQ) technologies \cite{preskill2018NISQ} have accelerated research into their practical applications. A key challenge is the efficient estimation of observables' expectation values for quantum states prepared by NISQ devices. This task forms the bedrock of many variational quantum algorithms \cite{cerezo2021variational,bharti2022noisy}, which hold potential for applications across diverse fields, including chemistry, materials science, and optimization. Within the realm of the Variational Quantum Eigensolver (VQE) \cite{peruzzo2014variational,tilly2022variational}, efficient measurement strategies are particularly crucial for tackling the electronic structure problem. The complexity of directly measuring the entire Hamiltonian is further compounded by its quantum nature, featuring a non-separable ground state with non-commuting terms \cite{bravyi2002fermionic,seeley2012bravyi,tranter2018comparison}. 

One method to address this challenge involves simultaneously measuring qubit-wise commuting (QWC) observables \cite{verteletskyi2020measurement,yen2020measuring,izmaylov2019unitary,gokhale2019minimizing}. QWC groups of observables can be found by applying the Minimum Clique Cover (MCC) \cite{karp1972reducibility} solver to the process of grouping Pauli operators by identifying the minimal number of cliques within a graph representation of QWC Hamiltonian terms. While the MCC problem is NP-hard, by applying heuristic algorithms to the MCC problem \cite{verteletskyi2020measurement}, simultaneous measurements of QWC observables greatly improves the efficiency of energy estimation.

Classical shadows \cite{classical-shadows} present an alternative method for energy estimation. With a quantum state $\rho$ prepared on quantum hardware and a collection of $L$ $n$-qubit Pauli operators, the goal is to accurately estimate their expectation values with minimal resources. The classical shadow protocol performs uniformly random Pauli-basis measurements on all qubits and then post-processes these measurements to estimate the expectation values of local observables. This method allows for the simultaneous estimation of numerous local observables. However, when the task is to estimate a specific set of target observables, uniform sampling in a randomized measurement scheme is suboptimal. To address the substantial inefficiency of the random sampling procedure in classical shadows, several methods have been proposed, which add a preprocessing stage. The derandomized shadows approach~\cite{derandomized-shadows} offers a deterministic protocol tailored to target observables, replacing the original randomized one. Subsequently, improved algorithms~\cite{locally-biased,hadfield2021adaptive,shadow-grouping,Wu_2023,Shlosberg_2023} that produce deterministic measurement bases before the quantum measurements, a process called the \textit{pre-processing} stage~\cite{dutt2023practical} in the classical shadow protocol, have been introduced. These methods reduce quantum resource requirements and enhance the precision of the randomized method at the cost of additional classical resources.

In this work, we propose \textit{Resource-Optimized Grouping Shadow} (ROGS), a novel approach for the efficient estimation of Pauli observables' expectation values on quantum states. Similar to the previous works~\cite{derandomized-shadows,locally-biased,hadfield2021adaptive,shadow-grouping,Wu_2023,Shlosberg_2023}, ROGS also acts as a pre-processing stage for the classical shadow protocol\footnote{Although we refer to our method as a shadow method, the techniques introduced here have potential applications beyond classical shadows. See more discussions in \Cref{sec: conclusion}.}. We consider efficiency from two aspects: the number of measurements ($n_{\mathrm{shot}}$) and the number of circuits ($n_{\mathrm{circuit}}$) prepared for measurements. {In the context of this paper, a \emph{circuit} refers to a specific sequence of quantum gates applied to qubits, followed by measurements in the computational basis (the $z$-basis)\footnote{\hypertarget{footnote:local random Pauli measurements} The measurement circuits consist solely of single-qubit gates, specifically Pauli measurement gates. Adopting the notation of \cite{classical-shadows}, each Pauli measurement gate is composed of a single-qubit rotational gate followed by a measurement in the computational basis (z-basis). The rotational gate of the circuit is the tensor product of unitaries on each qubit, $U = \bigotimes_{i=1}^n U_i$, where $U_i$ is selected from the set $\{ H, S^{\dagger} H, I \}$. Here, $H$ denotes the Hadamard gate, $S^{\dagger}$ is the adjoint of the phase gate, and $I$ is the identity gate. This combination of local rotational gates and z-basis measurements effectively realizes Pauli-basis measurements, enabling the estimation of expectation values for Pauli observables.
}. A \emph{shot} or \emph{measurement} refers to a single execution of this circuit on the quantum hardware, yielding a bitstring ($b \in\{\pm 1\}^{\otimes n}$) outcome. When estimating non-QWC observables, we need to measure multiple distinct circuits, each corresponding to a different Pauli operator. Here we refer to the number of measurements as the number of total shots we are allowed to gather from the quantum device or the number of times the device is queried; the measurement circuits are referred to as the distinct circuits used in the measurement process. 

We consider the task of estimating the expectation value of a Pauli Hamiltonian $${\mathcal{H}} = \sum_{\ell}a_{\ell}O_{\ell},$$ where $a_{\ell}\in \mathbb{R}$ is the coefficient and $O_{\ell}$ are Pauli operators, with a total budget of $M$ shots. We seek to allocate shots for each QWC group in the Hamiltonian optimally to increase the accuracy of energy estimation.

Our approach introduces a confidence bound for the error of the Hamiltonian, distinguishing it from the classical shadow protocol that focuses on bounding Pauli observables. This confidence bound serves as a cost function for optimizing the allocation of measurement resources among operator groups. When quantum resources are limited, minimizing the cost function concentrates measurements on groups of higher importance to energy estimation (i.e., groups containing a large number of observables or observables with a large sum of absolute coefficients), potentially causing the confidence bound to exceed one\footnote{Detailed in \Cref{subsec:shots allocation}}. However, the confidence bound acts as a guide for optimizing the measurement allocation, prioritizing the overall accuracy of the Hamiltonian estimation rather than providing a strict error guarantee for individual observables. The optimization of the measurement resource allocation based on the confidence bound is detailed in Algorithm \ref{alg:grouping-rogs-fixed}.

While previous works mostly focus on minimizing the number of measurements for energy estimation, in practice, the number of circuits is also a critical cost factor for the overall runtime, which has been largely overlooked~\cite{dutt2023practical}. The key distinction lies in the execution time associated with executing multiple shots of the same circuit versus executing single shots of many different circuits. From a hardware perspective, there is significant overhead in compiling and loading each unique circuit into the controller of the quantum device. This overhead is incurred for each distinct circuit, regardless of the number of shots. Consequently, executing a large number of shots for a single circuit is often much faster than executing the same total number of shots spread across many different circuits \cite{vazquez2024scaling, Tornow_2022}. Compared to previous methods, our approach achieves the lowest estimation error for a given number of measurements and uses significantly fewer circuits.

ROGS operates through a three-stage process, each of which is designed to enhance precision and reduce computational overhead:

\begin{enumerate}
[leftmargin=*,align=left,noitemsep,nolistsep]
\item \textbf{Max-Min Grouping of Pauli Observables}: We begin by applying a Minimum Clique Cover (MCC) algorithm to the QWC graph of Pauli observables in the Hamiltonian, which partitions the observables into \textit{disjoint} groups. Because certain observables can be included into distinct groups, we further expand each group to include the maximal number of additional observables, ensuring that observables within the expanded group still qubit-wise commute with each other. We map this group expansion problem to a \textit{Maximum Clique} problem \cite{bomze1999maximum}, which can be solved using various graph algorithms. This method allows observables within the same group to be simultaneously estimated using an identical set of Pauli-basis measurements. By minimizing the number of groups while maximizing the size of each group, we refer to this stage as Max-Min Grouping.

\item \textbf{Measurement Resource Allocation via Convex Optimization}: We resolve the challenge of distributing measurement resources among groups by defining it as a convex optimization problem, using a fixed number of measurements. The objective function, based on the confidence bound from mean estimation techniques like Hoeffding's inequality, sets an upper limit on the probability that the sample mean significantly deviates from the true mean (\Cref{subsec:various conf-bounds}). Employing a convex optimization solver to minimize this probability ensures that the allocation of measurement resources optimally minimizes the risk of estimation errors, given the predetermined level of measurement accuracy. This method guarantees that the resources are allocated in the most efficient way to minimize estimation errors within the constraints of the available measurements.

\item \textbf{Measurement and Estimation}: Having obtained the optimized measurement allocation for the groups of Pauli observables, we perform the corresponding Pauli-basis measurements on the quantum state of interest. Each group is measured using the allocated number of measurements determined by the convex optimization procedure above. Finally, we collect the measurement results and use them to estimate the expectation values of the target qubit Hamiltonian of Pauli observables.

\end{enumerate}

The remainder of the paper is organized as follows. 
In \Cref{sec:related works}, we remark on the key improvements of ROGS compared to previous works, focusing on proposed measurement frameworks for the quantum measurement stage of the classical shadow protocol.
In \Cref{sec:Algorithm Proposal}, we provide a detailed description of the ROGS method, presenting strategies for the optimal distribution of measurement resources across groups. Specifically, \Cref{subsec: conf-bound} offers theoretical assurances for the precision of our estimates by establishing tail-bounds on empirical estimators of the state's energy when employing grouping strategies. These bounds assist in assessing the accuracy and practicality of contemporary leading-edge measurement schemes. Additional details on refining the allocation of shots are discussed in \Cref{sec:fine grained shots allocation}, which focuses on the granular distribution of shots across groups. In \Cref{sec:exp}, we present numerical results benchmarking our molecular energy estimator against a series of increasingly complex molecules, comparing our findings to previous studies. This is followed by a discussion in \Cref{sec: discussion} on the statistical and physical interpretations of why our method outperformed others.  We conclude in \Cref{sec: conclusion}. The proof of our main theorem refers to \Cref{subsec:various conf-bounds}, as we have ignored some properties of the Hamiltonian in our main theorem, and supplementary confidence bounds are provided there. We discuss a mean estimator for heavy-tailed distributions, the median-of-means estimator, in \Cref{subsec:MoM}, which is an essential element of our methodology due to the generally high operator hit rate.\footnote{\hypertarget{footnote:hit rate}The hit rate 
is the number of measurements for which a Pauli measurement provides meaningful information about an observable following the definition in \cite{derandomized-shadows}. If the hit rate is zero, then the estimate of the target observable cannot be improved by measuring that Pauli.} 

\vspace{.1in}

\textbf{Notations.}~We adopt the notation $\{a_i\}_{1\leq i\leq n} := \{a_1,\dots, a_n\}$ to represent a set of objects indexed by $i$. We will use $\{a_i\}$ for simplicity if there is no confusion from the context. The symbol $|\cdot|$ signifies the number of terms in, or the dimension of, the object in question. Furthermore, $\hat{a}$ is used to denote a measurable function that serves as the mean estimator for the random variable $a$, derived from a sample $\{a_i\}$ of i.i.d.(independent and identically distributed) copies of $a$.

\section{Discussion of Prior Works}\label{sec:related works}

% In this section, we discuss some related works in detail and compare their methodological difference with ROGS.

Classical shadows \cite{classical-shadows} offers an efficient approach to predicting properties of quantum states using exponentially fewer measurements. Given a quantum state, $\rho$, prepared on quantum hardware and a set of Pauli operators, the classical shadow protocol aims to efficiently estimate their expectation values by performing uniformly random Pauli-basis measurements on all qubits and post-processing the measurement outcomes. By applying random unitaries $U$ (defined in footnote~\hyperlink{footnote:local random Pauli measurements}{2}) to rotate the state $\rho \rightarrow U^\dagger \rho U$, measuring in the computational ($z$-)basis, and storing classical snapshots $U^\dagger |\hat{b}\rangle\langle\hat{b}| U$, one can construct a classical shadow $\hat{\rho}$ of the quantum state:
\begin{align}\label{eqn:random unitary}
\hat{\rho} = \mathcal{M}^{-1}\left(U^\dagger|\hat{b}\rangle\langle\hat{b}|U\right),
\end{align}
where $\mathcal{M}$ is a quantum channel determined by the unitary ensemble. Repeating this process $N$ times yields a set of classical shadows, i.e., the union of all classical snapshots
\begin{align}\label{eqn:classical shadow}
    S(\rho; N) = \{\hat{\rho}_1, \dots, \hat{\rho}_N\},
\end{align}
which can be used to estimate expectation values, $\hat{O}_\ell = \mathrm{Tr}(O_\ell\hat{\rho})$, of observable $O_\ell\in \{O_\ell\}$.
    
While the classical shadow protocol allows for the simultaneous estimation of many local observables, it is suboptimal when targeting a specific set of observables. To improve upon the random sampling procedure in classical shadows, several pre-processing techniques have been proposed. These techniques aim to replace the random unitaries used to generate the classical shadow (\ref{eqn:classical shadow}) with fixed unitaries tailored to the target observables, thereby enhancing the efficiency and accuracy of the estimation process. The Derandomized Shadow approach \cite{derandomized-shadows} starts with the randomized classical shadow protocol and iteratively replaces random single-qubit Pauli measurements with deterministic ones, ensuring that the new partially derandomized protocol performs at least as well as the previous one. This process continues until all randomized measurements are replaced with a deterministic protocol tailored to the target observables. Similarly, importance sampling is employed in \cite{locally-biased,hadfield2021adaptive} to strategically select measurement bases and reduce estimation error, particularly for larger quantum systems. The work in \cite{shadow-grouping} further enhances derandomized shadows by exploiting qubit-wise commutation among the Pauli observables of interest. Building on these ideas, the Overlapped Grouping Measurement (OGM) framework \cite{Wu_2023} combines the advantages of importance sampling, observable compatibility, and classical shadows by arranging measurements into overlapped groups of compatible observables. Shlosberg et al. \cite{Shlosberg_2023} propose an adaptive algorithm, AEQuO, to efficiently estimate quantum observables represented as sums of Pauli operators. AEQuO uses overlapping groups of Pauli strings, considers general commutation relations, and adjusts measurements based on prior results. It applies a `bucket-filling' approach and a machine learning method to estimate both the mean and variance, cutting down on errors through post-processing.

While all previous methods aim to derandomize the classical shadow protocol, ROGS introduces a novel Max-Min grouping strategy. Instead of iteratively searching for a local minimum, ROGS employs convex optimization to determine the globally optimal resource allocation. Furthermore, ROGS explicitly aims to minimize the number of unique measurement circuits ($n_{\mathrm{circuits}}$), which is a key distinction from other methods.

Notice that in the classical shadow protocol, the error bounds are established for each observable, and the overall error bound is derived by applying the union bound. In contrast, our approach introduces a confidence bound that directly bounds the error of the Hamiltonian, which is the weighted sum of all observables. This confidence bound serves as a cost function for optimizing the allocation of measurement resources among operator groups.

By minimizing the confidence bound, we aim to achieve the best possible estimation accuracy for the Hamiltonian given the available quantum resources, even if the bound itself may not be tight.

\section{Algorithm Proposal}\label{sec:Algorithm Proposal}
In this section, we provide a detailed exposition of our ROGS method, which encompasses the strategy for overlapping grouping and the methodology for allocating quantum resources among these groups to attain optimal estimation results.

\subsection{Max-Min Grouping Protocol}\label{subsec:grouping}

Consider a qubit system of size $n$. The set of Pauli operators consists of tensor products of local Pauli operators $(O_\ell)_{\ell\leq L}$. These operators can be organized into groups by qubit-wise commutation (QWC), with each group comprising operators that qubit-wise commute with each other, i.e. Pauli operators $O_\ell=\bigotimes_{i=1}^n P_i^{(\ell)}$ are QWC,
\begin{align}
\begin{split}
   [O_\ell, O_r]_{\mathrm{QWC}} =0~~  \mathrm{iff} ~~[P_i^{(\ell)},P_i^{(r)}&]=0,\nonumber\\
&\forall~~ 1\leq i \leq n.
\end{split}
\end{align}
Previous studies, such as those by Verteletskyi et al.~\cite{verteletskyi2020measurement}, have concentrated on identifying the smallest possible number of these mutually non-commuting groups, a task that is generally NP-hard. This approach represents the QWC relationships between Hamiltonian terms as a graph, where the nodes represent the Pauli operators and the edges denote QWC relations between the Pauli operators. A group of QWC Pauli operators corresponds to a clique in the graph (nodes that are fully connected by edges). Thus, the challenge of operator grouping is reformulated as the problem of finding cliques in the QWC graph. Hence, finding the optimal grouping of the Pauli operators is equivalent to the \textit{Minimal Clique Cover} (MCC) problem, which aims to partition the graph into the fewest number of fully connected sub-graphs (cliques, i.e., \textit{non-overlapping groups} of Pauli observables), where each clique represents a group of QWC terms that can be measured together with the same circuit. Although the MCC problem is NP-hard in general, several polynomial-time heuristic algorithms, such as greedy coloring, have been explored in \cite{verteletskyi2020measurement} to find approximate solutions. Since the MCC strategy is an important step in our algorithm for reducing the number of measurement circuits, we have illustrated the MCC with a toy model in the left panel of Fig.~\ref{fig:maxmin_group}.

Denote $\{\mathcal{C}_\alpha\}$ as a set of disjoint Pauli-operator groups identified by MCC. These groups $\{\mathcal{C}_\alpha\}$ are directly associated with the measurement gates employed in classical shadows, represented by $\mathcal{P}_\alpha = P_1^{\alpha}\otimes\dots\otimes P_n^\alpha$, $P_i\in\{X,Y,Z\}$. The MCC approach identifies non-overlapping groups of observables, however, we note that the non-overlapping requirement is unnecessary \cite{Wu_2023}. By removing this requirement, we can expand the groups to further improve the measurement efficiency. This expansion is achieved by maximizing the overlaps between these originally disjoint groups. Specifically, for each group $\mathcal C_\alpha$, we expand it by including additional Pauli operators that (i) were originally not in $\mathcal C_\alpha$, (ii) QWC with every operator in $\mathcal C_\alpha$, and (iii) are QWC with each other. From a graph theory perspective, we lift the constraint that cliques are disjoint and then maximize the size of each clique while maintaining the same number of cliques as MCC. This procedure is illustrated in Fig~\ref{fig:maxmin_group}. We summarize our max-min grouping procedure below, which is detailed in \Cref{alg:MM-grouping}:
\onecolumn

\vspace{0.2in}
%\begin{algorithm}
{
\begin{itemize}[leftmargin=*,align=left,noitemsep,nolistsep]

   \item[]\textbf{Given:}  A graph $G = (\{O_\ell\}, \mathcal{Q}_c)$ is defined, with $\{O_\ell\}$ representing the set of Pauli operators and $\mathcal{Q}_c$ indicating the QWC relationships between them, i.e. for $O_i,O_j\in\{O_\ell\}$ qubit-wisely commute with each other if $(O_i,O_j)\in \mathcal{Q}_c$, i.e. $[Q_i,Q_j]_{\mathrm{QWC}}=0$.
\vspace{0.1 in}
\item[]\textbf{Minimize:}\label{minimize} The objective is to minimize the number of groups $|\{\mathcal{C_\alpha}\}|$ such that the union of all groups $\cup_{\alpha} \mathcal{C}_\alpha$ fully covers the set of all Pauli operators in the Hamiltonian~\cite{verteletskyi2020measurement}. This must be achieved while ensuring that operators within each $\mathcal{C}_\alpha$ are QWC:
\begin{align}\label{eqn:min-clique cover}
   \{\mathcal{C}_\alpha\} =\left\{ \arg\min_{\{\mathcal{C}_\alpha\}}|\{\mathcal{C}_\alpha\}|\Bigg\lvert[O_i,O_j]_{\mathrm{QWC}}=0,~\forall O_i,O_j\in \mathcal{C}_\alpha,~~~\cup_{\alpha} \mathcal{C}_\alpha = \{O_\ell\}_{\ell\leq L},~\mathcal{C}_\alpha\cap \mathcal{C}_\beta = \emptyset,~\forall \alpha\neq \beta\right\}
\end{align}
\item[]\textbf{Maximize:} \label{maximize} The next step involves maximizing the overlap between cliques for each $\mathcal{C}_\alpha$ by including the maximum clique of all Pauli operators that are QWC with every operator in $\mathcal{C}_\alpha$ and with each other. This maximization can be defined as:
\begin{align}\label{eqn:max-clique}
&\mathcal{C}_{\alpha}^{\mathrm{add}} =\left\{\max_{\mathcal{C}_{\alpha}^{\mathrm{add}}\subseteq \{O_\ell\}\backslash\mathcal{C}_{\alpha}}|\mathcal{C}_{\alpha}^{\mathrm{add}}| \bigg\lvert [O_i,O_j]_{\mathrm{QWC}},~ [O_j,O_k]_{QWC} =0,~~~\forall O_i\in \mathcal{C}_\alpha,~\forall O_j, O_k\in \mathcal{C}_{\alpha}^{\mathrm{add}}\right\}
\end{align}
The expanded group\footnote{We will henceforth call the expanded clique $\mathcal{C}_{\alpha}^{\mathrm{MM}}$, obtained by the above procedure, the overlapped group.} obtained by the above overlapped grouping procedure, denoted as $\mathcal{C}_\alpha^{\mathrm{MM} }$, is the union of the original group $\mathcal{C}_\alpha$ and the added operators $\mathcal{C}_\alpha^{\mathrm{add}}$, which QWC with all operators in $\mathcal{C}_\alpha^{\mathrm{add}}$. This can be expressed as: 
\begin{align}\label{eqn.overlapped grouping}
    \mathcal{C}_\alpha^{\mathrm{MM} }= \mathcal{C}_\alpha\cup \mathcal{C}_{\alpha}^{\mathrm{add}}.
\end{align}
\end{itemize} }

\begin{figure}[tbp!]
    \centering    \includegraphics[width=1.0\textwidth]{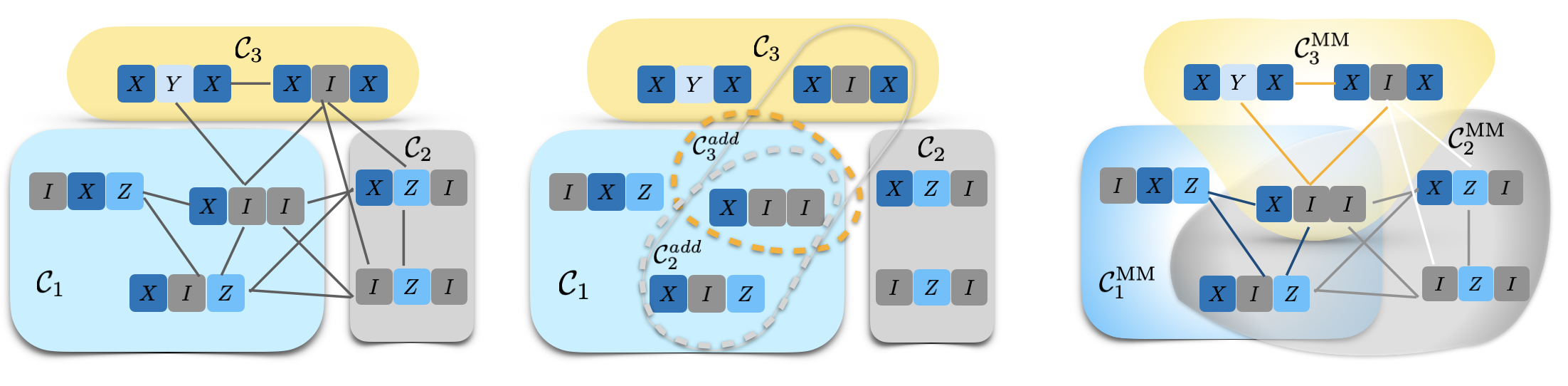}
    \caption{\textit{A graphical illustration of our Max-Min grouping protocol.} Each vertex, which is depicted as a Pauli string, represents a Pauli observable, with individual Pauli operators denoted by squares of varying colors.  \textit{Left} [\Cref{eqn:min-clique cover}: Minimize]: solid lines connect vertices that commute with each other. One of the minimum clique covers is depicted, with operators within the same shadows belonging to the same group $(\mathcal{C}_\alpha)_{\alpha = 1,2,3}$. \textit{Middle} [\Cref{eqn:max-clique}: Maximize]: the process of maximizing overlaps between groups to obtain $\mathcal{C}^{\mathrm{MM}}_\alpha$ is shown. In this example, operators inside the solid gray line are all operators that commute with $\mathcal{C}_2$, and the maximum clique of those operators that commute with each other (inside the gray dashed line) is the $\mathcal{C}_2^{\mathrm{add}}$ we are looking for to include into $\mathcal{C}_2$, and so it is with $\mathcal{C}_3^{\mathrm{add}}$ (circled by the yellow dashed line). \textit{Right} [\Cref{eqn.overlapped grouping}]: the overlapped shaded region $\mathcal{C}_2^{\mathrm{add}}$ includes operators in $\mathcal{C}_1$ that also commute with all operators in $\mathcal{C}_1^{\mathrm{MM}} = \mathcal{C}_1$ and $\mathcal{C}_{2,3}^{\mathrm{MM}} = \mathcal{C}_{2,3} \cup \mathcal{C}_{2,3}^{\mathrm{add}}$. The Pauli gates associated with group $\mathcal{C}_{1,2,3}$ are $\mathcal{P}_{1}=X_1\otimes X_2 \otimes Z_3$, $\mathcal{P}_2 = X_1\otimes Z_2\otimes I_3$ and $\mathcal{P}_3 = X_1\otimes Y_2\otimes X_3$; however, with the maximization procedure considered, measurement gates for $\mathcal{C}^{\mathrm{MM}}_{2}$ will have a fixed measurement gate for the third qubit and changes to $\mathcal{P}_2 = X_1 \otimes Z_2 \otimes Z_3$.}
    \label{fig:maxmin_group}
\end{figure}

\twocolumn

\begin{algorithm}[t!]
\small
\caption{ {\footnotesize \textbf{(MaxMin Grouping)} }}\label{alg:MM-grouping}
\begin{algorithmic}[1]
\Require
Qubit Hamiltonian ${\mathcal{H}} = \sum_{\ell}a_{\ell}O_\ell$ with all terms $O_\ell$ Pauli operator.

\Ensure 
MaxMin groups $\{\mathcal{C}^{\mathrm{MM}}_\alpha\}$ of all terms $\{O_\ell\}$ in ${\mathcal{H}}$.
\vspace{.1in}

\State \textbf{QWC Graph}: Construct the QWC graph $G = (\{O_l\}, \mathcal Q_c)$

\State \textbf{Minimize}: Apply a minimum clique cover solver~\cite{verteletskyi2020measurement} to $G$ output $\{\mathcal{C}_\alpha\}_{1\leq \alpha\leq A}$

\For{$\alpha=1$ to $A$} \Comment loop of over groups $\mathcal{C}_\alpha$%, with $\mathcal{P}_\alpha$ the measurement gates associate with $\mathcal{C}_\alpha$  
    \State initialize $\mathcal{C}_\alpha^{\mathrm{QWC}}=[~]$, 
    \While{$O_\ell\notin \mathcal{C}_\alpha$} \Comment loop over $O_\ell\notin \mathcal{C}_\alpha$
        \If{$O_\ell$ qubit-wise commutes with $\mathcal{P}_\alpha$ } \State append $O_\ell$ to $\mathcal{C}_\alpha^{\mathrm{QWC}}$
        \EndIf
    \EndWhile
\State Apply a maximum clique solver~\cite{boppana1992approximating} to $\mathcal{C}_\alpha^{\mathrm{QWC}}$, output  $\mathcal{C}_\alpha^{\mathrm{add}}$.% find the maximum clique $\mathcal{C}_\alpha^{\mathrm{add}}$ (that QWC with each other)
\State \textbf{Maximize}: $\mathcal{C}_\alpha^{\mathrm{MM}} = \mathcal{C}_\alpha.\mathrm{Append}[\mathcal{C}_\alpha^{\mathrm{add}}]$%Append $\mathcal{C}_\alpha^{\mathrm{add}}$ to $\mathcal{C}_\alpha$ to get maximum overlapped group $\mathcal{C}_\alpha^{\mathrm{MM}}$
\EndFor
\end{algorithmic}
\end{algorithm}

The maximize process mainly helps us decide the measurement gates of qubits that have not been assigned by the minimize procedure. For example, in Fig.~\ref{fig:maxmin_group}, the Pauli gates associated with $\mathcal{C}_2$ is changed from $X_1\otimes Z_2\otimes I_3$ to $X_1 \otimes Z_2 \otimes Z_3$ upon including $\mathcal{C}_2^{\mathrm{add}}$. This max-min grouping procedure ensures full utilization of each measurement. This maximization step is realized in~\Cref{alg:MM-grouping} through steps $3$ to $12$.
When we perform measurement for the group $\alpha$, using the measurement gates $\mathcal{P}_\alpha$, an operator will be measured if it belongs to the group $\alpha$. The overlap maximization step in our approach can increase the hit rates of observables that belong to the overlaps between operator groups. By maximizing the overlap, we ensure that the observables are measured more frequently within a fixed measurement budget, thereby improving the overall estimation accuracy. In general, this procedure guarantees high hit rates on observables, which is crucial for achieving precise estimates of the target quantities.

In the Overlapped Grouping Measurement (OGM) framework~\cite{Wu_2023}, as well as in other grouping schemes, the central task is to find compatible sets of Pauli operators for simultaneous measurement. While our Max–Min grouping strategy is guided by the Minimum Clique Cover (MCC) heuristic (see discussion above), OGM and other related methods (e.g.,~\cite{Shlosberg_2023}) often adopt a simpler \textit{largest degree first} (LDF) rule or other greedy approaches. Generally, LDF aims to reduce the problem complexity by iteratively placing observables with the highest graph degree (i.e., the most edges in a compatibility graph) into existing or newly created cliques. This scheme often provides a decent partition without fully solving the MCC problem (which is NP-hard), but it can be suboptimal in cases where local high‐degree choices do not translate into overall optimal grouping. In contrast, heuristics inspired by MCC may identify larger compatible sets and thus yield fewer final groups in practice, depending on the specific algorithm used. On the other hand, LDF and similar degree‐based algorithms tend to be faster to implement in large‐scale settings, at the risk of increasing the number of overall groups. In practice, which approach is preferable often depends on the trade‐off between computational overhead for determining groups and the downstream cost of executing distinct measurement circuits. Our MCC‐based Max–Min grouping leverages the clique cover perspective to minimize the circuit count while still accommodating overlaps, thereby giving both a systematic criterion for forming QWC groups and—once expanded—maximizing overlap among them. This ensures efficient shot usage and avoids redundant basis measurements, which can be crucial in resource-limited scenarios.

The next task is to allocate measurement resources to each group to minimize the error.

\subsection{Confidence Bound}\label{subsec: conf-bound}

Upon identifying the optimal operator grouping, our objective shifts toward determining the optimal allocation of measurement shots per group. Heuristically, assigning measurement budgets evenly across all the groups may seem reasonable, however, because a quantum Hamiltonian is a \emph{weighted} average of Pauli observables, different operators would have varying importance to the accuracy of the energy estimation. Hence, allocating varying numbers of shots for different operator groups could potentially lead to more accurate estimation. Consequently, a cost function is imperative for guiding the allocation of shots to each group of observables. 

Given the number of measurement shots $M$, we aim to achieve the most efficient group-wise measurement allocation, $\{M_\alpha\}$ with $\sum_\alpha M_\alpha = M$, minimizing the overall estimation error of the target Hamiltonian.

Consider the mean estimator for a Hamiltonian $\hat{{\mathcal{H}}} :=\sum_\ell a_\ell\hat{O}_\ell$, where $\hat{O}_\ell$ is the expectation value estimator for $\langle O_\ell\rangle_\rho = \mathrm{Tr}(\rho O_\ell)$. We group the operators into QWC max-min groups $\{\mathcal{C}^{\mathrm{MM}}_\alpha\}_{1\leq \alpha\leq A}$. Our goal is to find the optimal allocation of measurement shots per group, denoted by $w_\alpha$. The expectation value of the Hamiltonian is then estimated by averaging the corresponding measurement outcomes. In practice, we use the median-of-means estimator (see Eq.~\ref{eqn.mom estimator}) as the mean estimator for $\langle {\mathcal{H}}\rangle_\rho$, with details provided in \Cref{subsec:MoM}. The following theorem introduces a confidence bound that limits the error in the Hamiltonian expectation value for shots allocated to the max-min groups.

\begin{theorem}{(Confidence bound)}\label{theorem:Confidence bound}
    Suppose that a set of group $\{\mathcal{C}^{\mathrm{MM}}_\alpha\}_{1\leq \alpha\leq A}$ given as a set of max-min groups $\{C_\alpha^{\mathrm{MM}}\}$ of a set of Pauli observables $\{O_\ell\}_{\ell\leq L}$ of Hamiltonian ${\mathcal{H}} = \sum_\ell a_\ell O_\ell$, and a list assigns measurements to each group $\{w_\alpha\}$, and accuracy parameters $\epsilon,\delta$, s.t.
    \begin{align}\label{conf-bound}
    &\mathrm{Conf}\left[\epsilon,\{w_\alpha\};M,\{C_\alpha^{\mathrm{MM}}\}\right] :=\nonumber\\
    & 2\sum_{\ell=1}^{L} \exp\left(-\frac{\epsilon^2 M}{2(\sum_\ell|a_\ell|)^2}\sum_{\alpha =1}^A\mathrm{idx}_{\ell,\alpha}\;w_\alpha  \right)= \delta
\end{align}
which subjected to $\sum_\alpha w_\alpha = 1$. 

The indicator $\mathrm{idx}_{\ell,\alpha}$ is define as
    \begin{align}
\mathrm{idx}_{\ell,\alpha}=\left\{
        \begin{array}{ll}
        1 ~~~~~& \text{if } O_\ell \in \mathcal{C}^{\mathrm{MM}}_\alpha\\
         0 & \text{otherwise} 
        \end{array}\right.
    \end{align}

    Then, a collection of $M$ independent classical shadows allows for accurately predicting the expectation Hamiltonian for the given quantum state $\rho$:
    \begin{align}\label{eqn:The:error bound of energy estimation}
        \left| \hat{{\mathcal{H}}} -\langle{\mathcal{H}}\rangle_{\rho}\right|\leq \epsilon
    \end{align}
     with probability $1-\delta$.
\end{theorem}

The detailed proof is provided in Appendix \ref{subsec:various conf-bounds}. It should be highlighted that our proof adopts a similar process to the original classical shadow protocol in measuring a set of operators. We establish error bounds for each operator, and the bound on the Hamiltonian, which is the weighted sum of all observables, is derived using the union bound over all operators. This approach is analogous to the classical shadow protocol, where the error bounds are derived for each individual observable, and the overall error bound is obtained by applying the union bound.

This confidence bound serves as an overall cost function for our measurement resource allocation, where $\epsilon$ represents the error bound of the Hamiltonian. When quantum resources are limited, minimizing the cost function leads to a concentration of measurements on a few groups, i.e., groups containing a large number of observables. In such cases, it is possible for the cost function to exceed one, causing the confidence bound to break down. However, this is acceptable since the confidence bound acts as a cost function for shot allocation under limited quantum resources, and it is crucial to recognize that in these situations, accuracy is not guaranteed. The confidence bound is used as a guide for optimizing the measurement allocation, rather than as a strict guarantee on the estimation error.

In most scenarios, the confidence bound provided in \Cref{theorem:Confidence bound} offers an effective guide for optimizing the allocation of measurement resources across groups, prioritizing the overall accuracy of the Hamiltonian estimation. However, in certain extreme cases, such as when a Hamiltonian contains groups with significantly imbalanced sizes and coefficient magnitudes (see \Cref{subsection:conter exp}), the confidence bound in \Cref{theorem:Confidence bound} may not provide the optimal allocation\footnote{Although \Cref{theorem:Confidence bound} does not yield optimal bounds without sampling in the hyperparameter space, it will produce accurate values if we follow the process detailed in \Cref{alg:coarse-graining ROGS}.}. In such cases, the supplementary confidence bound introduced in \Cref{subsec:group Hoeffding Bound} can be more useful. This alternative bound takes into account the coefficients of the observables within each group, allowing for a more nuanced allocation of measurement resources. By considering each group's relative importance based on its size and the magnitudes of its observable coefficients, the supplementary confidence bound can help mitigate the potential breakdown of the primary bound in these extreme scenarios. Nonetheless, for most practical applications, the confidence bound in \Cref{theorem:Confidence bound} remains a robust and reliable tool for guiding the measurement optimization process.

\subsection{Measurement Resource Allocation}\label{subsec:shots allocation}

Given the confidence bound proposed in~\Cref{theorem:Confidence bound}, which is used as a cost function, the next step involves solving the measurement resource allocation problem by minimizing the confidence bound, which is presented in Algorithm \ref{alg:grouping-rogs-fixed}. By optimizing the allocation of measurement resources based on the confidence bound, we aim to achieve the best possible estimation accuracy given the available quantum resources, even if the confidence bound itself may not always provide a tight error guarantee.

\begin{algorithm}[t!]
{\small
\begin{algorithmic}[1]
\caption{{\small \textbf{(ROGS)}}}
\label{alg:grouping-rogs-fixed}
\Require
Measurement budget $M$, accuracy parameter $\epsilon$, and Hamiltonian ${\mathcal{H}} = \sum_{\ell}a_{\ell}O_\ell$. 

\Ensure
A fixed Pauli basis measurement recipe $\{\{\mathcal{P}^{(m)}_\alpha\}_{1\leq m\leq M_\alpha}\}\in\{X,Y,Z\}^{n\times M}$ for the classical shadow quantum measurement stage. 

\vspace{.1in}

\State Identify the max-min groups $\{C_\alpha^{\mathrm{MM}}\}$ of operators $\{O_\ell\}$ in ${\mathcal{H}}$ using Algorithm~\ref{alg:MM-grouping}.
\State Apply a convex optimization solver to find the optimal distribution $\{w_\alpha\}$ of measurements among groups to minimize the confidence bound,
\begin{equation}\label{eqn.argmin-distribution}
    \{w_\alpha^*\} = \arg\min_{\{w_\alpha\}} \mathrm{Conf}\left[\epsilon,\{w_\alpha\};M,\{C_\alpha^{\mathrm{MM}}\}\right].
\end{equation}
\State Obtain measurement recipes $\{\{\mathcal{P}^{(m)}_\alpha\}_{1\leq m\leq M_\alpha}\}$ which include $\{X,Y,Z\}^{n\times M}$ based on $\{M_\alpha\}:=\{[w_\alpha^*M]\}$ and one to one correspondence between groups $\{\mathcal{C}_\alpha^{\mathrm{MM}}\}$ and Pauli measurement gates $\{\mathcal{P}_\alpha\}$.
\end{algorithmic}
}
\end{algorithm}

Now, we can determine the allocation of the measurement budget among groups based on the confidence bound \eqref{conf-bound} derived in \Cref{theorem:Confidence bound}. In essence, this bound provides a way to quantify how the total estimation error probability depends on the distribution of the measurements (shots) among the groups. By transforming the confidence bound into a log-sum-exp function (see Eq.~\eqref{conf-bound}), we obtain a smooth and convex objective that naturally lends itself to efficient numerical minimization.

Concretely, we introduce continuous variables $\{w_\alpha\}$, each representing the fraction of the total measurement budget $M$ allocated to group $\mathcal{C}_\alpha^{\mathrm{MM}}$. We then solve~\cref{eqn.argmin-distribution}
where ``$\mathrm{Conf}[\dots]$'' refers to the log-sum-exp confidence bound and $\epsilon$ is an error parameter. The solution 
$\{w_\alpha^\ast\}$ provides an optimal \emph{relative} shot allocation across groups, in the sense that it achieves the smallest upper bound on the probability of exceeding the target estimation error $\epsilon$. Since $\mathrm{Conf}[\dots]$ has a known convex form, we can use standard convex optimization toolboxes such as CVXPY~\cite{cvxpy} to solve for $\{w_\alpha^\ast\}$ efficiently, even for relatively large problems.

In effect, the solver balances the measurement `importance' of each group. Groups with many high-coefficient or critical operators (in terms of their Hamiltonian contribution) reduce the overall error bound more significantly when they receive more shots. This phenomenon will be discussed with greater detail in \Cref{sec: discussion}.
Once the optimal fractions $\{w_\alpha^\ast\}$ are obtained, we translate them directly into integer numbers of shots $M_\alpha = \lfloor w_\alpha^\ast\,M \rfloor$. Each $M_\alpha$ is then used to measure the group $\mathcal{C}_\alpha^{\mathrm{MM}}$ repeatedly under the same Pauli measurement circuit $\mathcal{P}_\alpha$. The set of all such circuits and shot counts $\{\{\mathcal{P}_\alpha^{(m)}\}_{1\leq m \leq  M_\alpha}\}$, constitutes our \textit{measurement recipe}\footnote{$\{\mathcal{P}^{(m)}_\alpha\}_{1\leq m\leq M_\alpha}$ denotes the repetition of $\mathcal{P}_\alpha$ for $M_\alpha$ times, where $\mathcal{P}_\alpha = P_1^{\alpha}\otimes\dots\otimes P_n^{\alpha}$ ($P_i$ is a Pauli measurement gate on site-$i$) and is associated with a single Pauli measurement on the given state, yielding a single classical snapshot \cite{classical-shadows} $\hat{\rho}_\alpha$. The subscript $\alpha$ labels this classical snapshot because all operators $O_\ell$ in ${\mathcal{H}}$ that belong to $\mathcal{C}_\alpha^{\mathrm{MM}}$ produce a non-zero random variable $\mathrm{Tr}(\hat{\rho}_\alpha O_\ell)\in\{-1, 1\}$, which is used in predicting the expectation $\hat{O}_\ell = \mathbb{E}[\mathrm{Tr}(\hat{\rho}_\alpha O_\ell)]$. Conversely, if $O_\ell\notin \mathcal{C}_\alpha^{\mathrm{MM}}$, this snapshot cannot be used to evaluate the expectation i.e. $\mathrm{Tr}(\hat{\rho}_\alpha O_\ell)=0$ for all $O_\ell\notin\mathcal{C}_\alpha^{\mathrm{MM}}$.} for energy estimation. In this way, the total measurement budget $M$ is divided among distinct circuits to optimize the overall accuracy of our Hamiltonian estimator.

\section{Results}\label{sec:exp}

In this section, we focus on estimating the ground state energy of small molecules with various encoding schemes: Jordan-Wigner (JW) \cite{jordan1993paulische}, Bravyi-Kitaev (BK) \cite{bravyi2002fermionic}, and Parity (P) \cite{bravyi2002fermionic, seeley2012bravyi}. These schemes transform the fermionic Hamiltonian of a specified molecule into qubit Hamiltonians, represented as weighted sums of Pauli operators. 
We benchmark against the exact calculation of their ground state energy by diagonalization of the Hamiltonians.

\subsection{Implementation} We obtain the qubit Hamiltonians of benchmark molecules with various encodings via Qiskit \cite{qiskit} and diagonalize these Hamiltonians numerically to obtain the corresponding ground states using SciPy \cite{scipy}. For the Max-Min grouping stage, the minimum clique cover is obtained via Qiskit as well, and the maximum cliques are found by solvers from NetworkX \cite{networkx}. In the measurement resource allocation stage, we adopt the convex optimization solver from CVXPY \cite{cvxpy}. Finally, we perform circuit simulations with the PennyLane lightning simulator \cite{pennylane}.

\subsection{Results} 

In Table \ref{table:number of circuits}, we benchmark ROGS against other measurement techniques for various small molecules (\chem{H_2}, \chem{Li H}, \chem{Be H_2}, \chem{H_2 O}, \chem{N H_3}) using the three encoding schemes. Each algorithm was allocated an equal number of measurement shots to compute the average error $|\hat{{\mathcal{H}}}-\langle {\mathcal{H}}\rangle_{\rho}|$. Through ten simulations and the subsequent calculation of their root mean square deviation, it is evident that the ROGS technique results in a lower estimation error compared to other methods. Notably, the ROGS approach often exceeds the performance of alternative strategies while requiring significantly fewer measurement circuits (see Table \ref{table:number of circuits}), demonstrating its efficient allocation of selective measurement resources. A key observation is that, with limited measurement shots, molecules with a larger number of operators and a greater number of mutually non-commuting groups tend to require measuring the same or even fewer groups. This occurs because, under a limited budget, accuracy is higher when focusing on the estimation of groups with more operators rather than measuring all groups with very few shots. This phenomenon illustrates the bias-variance tradeoff \cite{james2013introduction}, which can be interpreted using the concept of mean-field theory as we consider the high-weighted groups as the mean-field subspace that captures the dominant interactions within the system, which is detailed in~\Cref{sec: discussion}. ROGS performance is similar to the recently proposed method, adaptive estimation of quantum observables (AEQuO) \cite{Shlosberg_2023}. This method also seeks to optimally allocate measurements to overlapping groups. However, AEQuO allocates shots to minimize the variance of the target observable, as opposed to ROGS which minimizes the confidence bound in Theorem~\ref{theorem:Confidence bound}.

\begin{table}[tbp]
\centering
    \renewcommand{\figurename}{Table}
     \resizebox{\columnwidth}{!}{
    \begin{tabular}{|l|c|c|c||c|c|c|c|c|}
\hline
\shortstack{Molecule ($L$)} & Enc. & $n_{\mathrm{groups}}$  & ROGS  & Derand.& APS & OGM& AEQuO\\
\hline
\multirow{3}{*}{{\chem{H_2}~(185)}}& JW & 46 & \textbf{6} & 159&217&37&40\\
 & P & 34 & \textbf{5} & 77&116&39&30\\
 & BK & 34 & \textbf{8} &66&114&33&26\\
\hline
\multirow{3}{*}{{\chem{Li H}~(631)}} & JW & 136 & \textbf{11}& 643&259&63&126\\
 & P & 165 & \textbf{41}& 534&177&78&126\\
 & BK & 211 & \textbf{2} & 661&237&57&136\\
\hline
\multirow{3}{*}{{\chem{Be H_2}~(666)}} & JW & 140 & \textbf{24}& 624&462&56&126\\
 & P & 177 & \textbf{10} &630&310&76&115\\
 & BK & 193 & \textbf{10}& 671&300&80&110\\
\hline
\multirow{3}{*}{{\chem{H_2 O}~(1086)}} & JW & 224 & 41 & 781&560&\textbf{32}&171\\
 & P & 260 & \textbf{3} & 774&415&57&174\\
 & BK & 303 & \textbf{5}& 786&319&48&172\\
\hline
\multirow{3}{*}{{\chem{N H_3}~(3057)}} & JW & 618 & \textbf{5} &962&776&39&287\\
 & P & 720 & \textbf{3} &956&568&79&298\\
 & BK & 729 & \textbf{7}& 987&623&60&280\\
 \hline
\end{tabular}}
\caption{\textit{Empirical circuit cost benchmark:} This table shows the comparison of the number of circuits $n_{\mathrm{circuit}}$ required for estimating the ground states using a measurement budget of 1,000 shots of benchmark molecules (each with $L$ Pauli operators) across different encoding schemes: JW, P, and BK. The column labeled $n_{\mathrm{groups}}$ indicates the total number of overlapping groups, while the $n_{\mathrm{circuit}}$ column shows the significantly reduced number of circuits achieved with the grouping shadow technique. The abbreviations used are as follows: ROGS, derandomized classical shadow (Derand.) \cite{derandomized-shadows}, Adaptive Pauli Shadow (APS) \cite{hadfield2021adaptive}, overlapped grouping measurement (OGM) \cite{Wu_2023}, and adaptive estimation of quantum observables (AEQuO) \cite{Shlosberg_2023}.}
\label{table:number of circuits}
\end{table}

\begin{table}[htbp]
\centering
    \renewcommand{\figurename}{Table}
    \resizebox{\columnwidth}{!}{
    \begin{tabular}{|l|c|c||c|c|c|c|}
    \hline
       Molecule ($E_{\mathrm{GS}}$) & $\,$Enc.$\,$&ROGS & Derand.   & APS& OGM & AEQuO\\
       \hline
       \multirow{3}{*}{\chem{H_2} $(-1.86)$} & JW&\textbf{0.03}& {0.06}  &  0.07 &0.05&\textbf{0.03}\\
        & P &\textbf{0.01}  &{0.03} & 0.05&0.06&{0.02}\\
       & BK &\textbf{0.02}& {0.06}& 0.08&0.05&\textbf{0.02}\\
       \hline
       \multirow{3}{*}{\chem{Li H} $(-8.91)$} & JW &\textbf{0.02} & {0.03}  & 0.04&0.04&\textbf{0.02}\\
       & P &\textbf{0.02}& {0.03} &0.05&0.04&\textbf{0.02}\\
       & BK &\textbf{0.01}& {0.04}& 0.07&0.04&{0.02}\\
       \hline
       \multirow{3}{*}{\chem{Be H_2} $(-19.04)$} & JW &\textbf{0.02}& {0.06} & {0.06}&0.07&{0.04}\\
       & P&\textbf{0.03} & {0.09} & {0.06}&0.08 &{0.04}\\
       & BK&\textbf{0.04}& {0.06} &{0.06}&0.06&\textbf{0.04}\\
       \hline
       \multirow{3}{*}{\chem{H_2 O} $(-83.60)$} & JW &\textbf{0.09}& {0.12} & {0.11}&0.13&\textbf{0.09}\\
       & P&{0.08}& {0.22} & {0.11}&0.20 &\textbf{0.07}\\
       & BK &\textbf{0.07}& {0.20} & {0.10}&0.15&0.10\\
       \hline
       \multirow{3}{*}{\chem{N H_3} $(-66.88)$} & JW&\textbf{0.09} & {0.18}  & 0.13&0.15&0.10\\
       & P &\textbf{0.09}& {0.21} & 0.14&0.12&0.12\\
       & BK &\textbf{0.08}& {0.12}  & 0.11&0.14&0.11\\
       \hline
    \end{tabular}}
 \caption{\textit{Empirical benchmark:} This table shows the average estimation error (RMSE) across various molecules, encodings, and measurement strategies. The RMSE is calculated by simulating the process 10 times, each with 1,000 measurements. The first column details the molecule, along with its ground state electronic energy expressed in Hartree units. }
\label{table: ground state estimation across mole}
\end{table}

In Figure~\ref{fig:GSE error}, we show the energy estimation with error bars for the molecule \chem{Li H} using the BK encoding scheme. While all techniques have reasonable accuracy, ROGS in particular has a much smaller variance due to the deterministic allocation of measurements with convex optimization. This is an example of the bias-variance tradeoff. In~\Cref{fig:resource eff}, we show the quantum resource efficiency of ROGS for ground state energy estimation of small molecules (\chem{H_2}, \chem{Li H}, \chem{Be H_2}) using the BK encoding. ROGS consistently achieves target accuracies with the fewest shots and unique circuits compared to the derandomized classical shadow and AEQuO methods. The efficiency advantage of ROGS becomes more pronounced as the number of qubits increases. Interestingly, ROGS requires fewer shots for \chem{Li H} than \chem{H_2}, despite the latter having a smaller Hilbert space. This is due to the highly localized mean-field subspace (detailed in~\Cref{sec: discussion}) and low one-shot variance of operator expectation estimation in \chem{LiH}. By minimizing both the number of measurements and distinct circuits, ROGS addresses the significant overhead in compiling and loading circuits on quantum hardware, enabling more efficient characterization of molecular systems.

\begin{figure}[tbp]
    \centering
    
    \includegraphics[width=\linewidth]{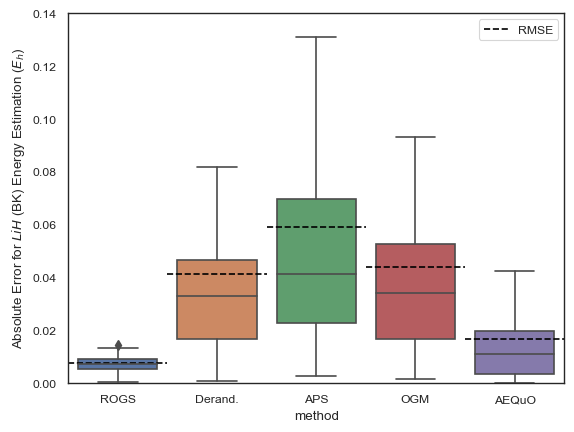}
    \caption{\small{\textit{Estimation Accuracy of Ground State Energy for \chem{LiH} with Bravyi-Kitaev Encoding \cite{bravyi2002fermionic} using various measurement strategies.}  This figure shows box plot of the absolute error of ground state energy estimations across 50 independent runs, each with 1,000 measurements. We compare ROGS with several methods:  Derandomized Classical Shadow (Derand.) \cite{derandomized-shadows}, Adaptive Pauli Shadow (APS) \cite{hadfield2021adaptive}, Overlapped Grouping Measurement (OGM) \cite{Wu_2023}, and adaptive estimation of quantum observables (AEQuO) \cite{Shlosberg_2023}. The solid line in the bar plot represents the mean, while the dashed line represents the root mean square error.} }
    \label{fig:GSE error}
\end{figure}

\begin{figure}[tbp]
    \centering
    \includegraphics[width=\linewidth]{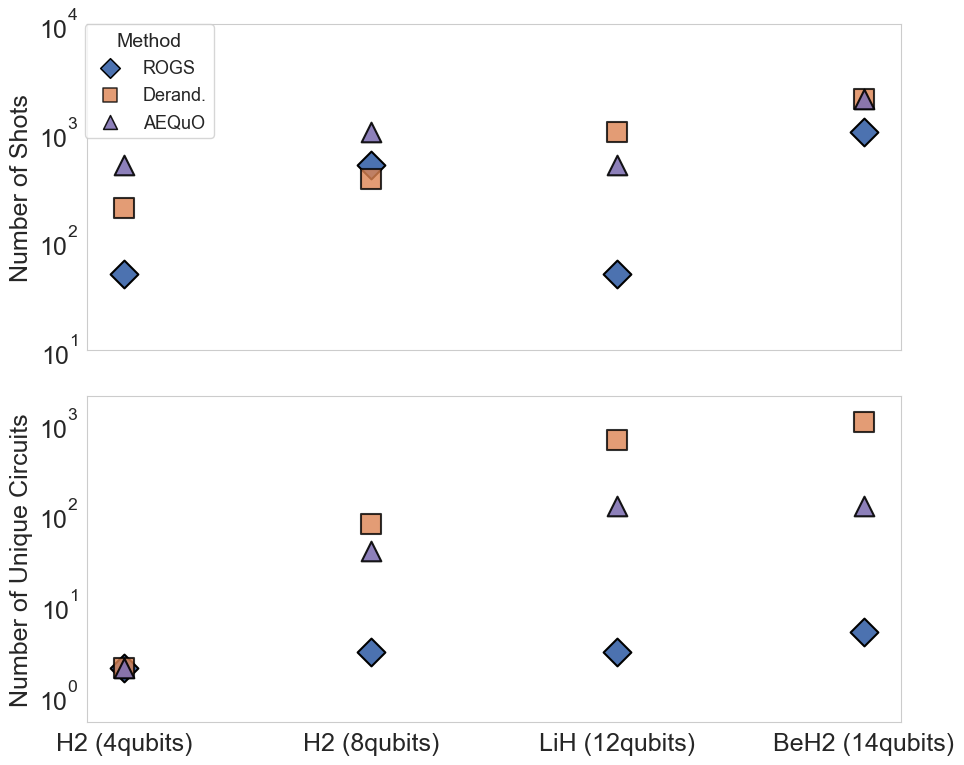}
\caption{
\small{ \textit{Quantum Resource Efficiency of Ground State Energy Estimation} for various molecules using the Bravyi-Kitaev encoding \cite{bravyi2002fermionic}. The upper plot shows the number of shots (quantum circuit executions), while the lower plot displays the number of unique quantum circuits required to reach a target accuracy of 0.03 Hartree ($E_h$) for different molecules and measurement strategies: ROGS (blue $\Diamond$), Derandomized Classical Shadow (Derand., orange $\square$), and adaptive estimation of quantum observables (AEQuO, purple $\triangle$).}
}
% 
%\textit{Quantum Resource Efficiency of Ground State Energy Estimation} for various molecules using the Bravyi-Kitaev encoding \cite{bravyi2002fermionic}. The upper plot shows the number of shots (quantum circuit executions), while the lower plot displays the number of unique quantum circuits required to reach target accuracies of 0.03 (yellow), 0.07 (green), and 0.1 (dark purple) Hartree ($E_h$) for different molecules and measurement strategies: ROGS ($\bigcirc$), Derandomized Classical Shadow (Derand., $\square$) \cite{derandomized-shadows}, and adaptive estimation of quantum observables (AEQuO, $\triangle$) \cite{Shlosberg_2023}. The color bar on the right indicates the error bound for the quantum resources required to achieve the corresponding accuracy.
    \label{fig:resource eff}
\end{figure}

As the provision of quantum resources increases, these resources tend to be distributed more evenly across all groups, as illustrated in Fig. \ref{fig:H2 shots allocation} using \chem{H_2} (encoding: BK, basis: 6-31g) as an example.  It is observed that the number of groups measured increases from six high-weighted groups to encompassing all groups. When the measurement resource is limited, the global optimum of the confidence bound concentrates resources on a subset of groups, which is beneficial due to the bias-variance tradeoff. This concentration of shots in high-importance groups diminishes as the available quantum resources increase. In the limit of infinite measurement shots, all observables must be estimated accurately to minimize the total estimation error. This is due to each group containing at least one unique Pauli observable that cannot be covered by the group maximization procedure. Consequently, as the number of shots approaches infinity, the allocation of shots across groups becomes more uniform to ensure that all observables are measured with enough resources. This behavior ensures that the estimation error of the Hamiltonian is minimized by accurately estimating all of its constituent observables, regardless of their distribution among the groups.

\begin{figure}[htbp]
    \centering
    \includegraphics[width=\linewidth]{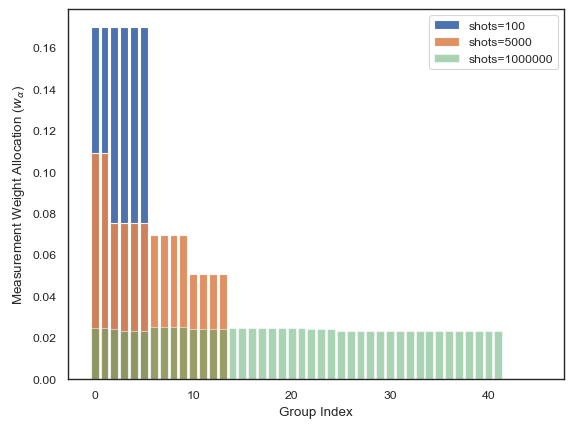}
    \caption{\small{Shot Allocation for \chem{H_2} Using Jordan-Wigner~\cite{jordan1993paulische} (JW) Encoding Across Varying Total Number of Experiments. This plot illustrates the allocation of shots as determined by convex optimization of the confidence bound (\Cref{theorem:Confidence bound}) for a specified $\epsilon$ over different total numbers of measurements $M$. It can be observed that with the total measurement shots $M$ increasing, the allocation of measurement weights across groups becomes more uniform. 
    }}
    \label{fig:H2 shots allocation}
\end{figure}

It is worth noting that the MCC problem is an NP-hard problem and is typically solved by applying heuristic algorithms \cite{verteletskyi2020measurement}, which do not guarantee to find the true minimum clique cover. The minimum clique cover found by \cite{verteletskyi2020measurement} for \chem{H_2} (encoding: BK, basis: 6-31g) results in $n_{\mathrm{groups}} = 46$ qubit-wise commuting groups (see~\Cref{table:number of circuits}). However, this number is over-counted, and in this example, four of the groups do not contain any unique Pauli operators, i.e., operators that belong only to a single group. In our shot allocation determined by \Cref{theorem:Confidence bound}, one can observe that no shots are allocated to these groups, even when more than enough quantum resources are provided (see \Cref{fig:H2 shots allocation}). This suggests that our approach based on \Cref{theorem:Confidence bound} could potentially provide a closer approximation to the optimal MCC solution by effectively identifying and excluding redundant groups that do not contribute much to the overall energy estimation.

\section{Discussion}\label{sec: discussion}
For the numerical experiments presented in \Cref{sec:exp}, it is found that excellent results can be achieved by measuring very few groups (fewer than the total number of groups, see Table \ref{table:number of circuits}). This is attributed to the fact that with a fixed measurement budget, omitting certain groups with insufficient measurements can significantly reduce the estimation variance at the cost of an increased bias in estimation. This tradeoff, widely recognized as the bias-variance tradeoff in the statistics community (see, for instance, \textit{Sec.~2.2.2} of Ref.~\cite{james2013introduction}), has been shown to be beneficial for mean estimation with finite samples.

Let us comment on the {\bf physical interpretation} of the ROGS method. From a physical standpoint, each group $\mathcal{C}_\alpha^{\mathrm{MM}}$ can be viewed as a mean-field subspace of the Hamiltonian \cite{izmaylov2019revising}\footnote{In the mean-field approach, the Hamiltonian is partitioned into a sum of mean-field Hamiltonians, each of which has eigenstates that are unentangled products of single-qubit states. This allows each mean-field Hamiltonian to be measured using a single set of single-qubit measurements in the appropriate single-qubit eigenbases. However, unlike in our context, the terms within each mean-field Hamiltonian \textit{do not necessarily} commute with each other.}. These subspaces are constructed by maximizing the number of mutually commuting operators within each group, essentially capturing the local mean-field interactions among the qubits. The mean-field approximation is often used to describe systems where the interactions between particles can be effectively averaged out, leading to a simplified description of the system. In the context of the overlapped grouping protocol (Sec.~\ref{subsec:grouping}), the larger groups with more mutually commuting operators can be seen as capturing these averaged interactions, providing a good approximation of the overall system behavior.

It is worth pointing out that the mean-field subspaces identified by ROGS not only correspond to groups with a large number of operators but also account for the magnitude of the coefficients associated with these operators. In the context of electronic structure Hamiltonians, the coefficients of Pauli operators can vary significantly, with some terms having much larger coefficients than others. As demonstrated in \cite{izmaylov2019revising}, these large coefficient terms often capture the dominant mean-field interactions within the system. Therefore, the high-weight groups in ROGS are those that contain either a large number of operators or operators with very large coefficients, or both\footnote{In cases where the Hamiltonian contains groups with significantly imbalanced sizes and coefficient magnitudes, the shot allocation given by the alternative confidence bound in \Cref{subsec:group Hoeffding Bound} may be more appropriate. This alternative bound takes into account the relative importance of each group based on both its size and the magnitudes of its observable coefficients, providing a more balanced allocation of measurement resources in such extreme scenarios.}.

When we have a limited number of measurements, allocating more resources to these mean-field subspaces allows us to better capture the dominant interactions within the system. By focusing on these larger groups, we are effectively prioritizing the measurement of the most significant contributions to the overall Hamiltonian. On the other hand, distributing the measurements evenly across all groups, including those with fewer operators, will lead to a less accurate representation of the system (see results in~\Cref{table:ROGS naive evenly distribution} \textit{Even distribution}). This is because the smaller groups may contain less significant interactions or may represent more local, "fluctuation" terms that deviate from the mean-field approximation.

By allocating more measurement resources to these mean-field subspaces, we are essentially focusing on the most relevant parts of the Hamiltonian that contribute to the overall system properties. This approach allows us to obtain excellent results with fewer measured groups, as the dominant interactions are well-captured within these larger mean-field subspaces. Hence the mean-field interpretation suggests that concentrating measurements on groups with more operators (i.e., larger mean-field subspaces) is an effective strategy when dealing with limited resources. In this approach, with limited measurement resource, we mainly the capture of the most significant interactions within the system, leading to accurate results while measuring fewer groups overall. The bias-variance tradeoff can be seen as a manifestation of this mean-field approximation, where focusing on the dominant mean-field subspaces reduces the variance in the measurement results at the cost of potentially introducing some bias by neglecting the smaller fluctuation terms.

\section{Conclusion and Outlook}\label{sec: conclusion}

In this work, we introduce a novel method, Resource-Optimized Grouping Shadow (ROGS), for efficiently estimating the expectation values of a set of Pauli observables on a quantum state. The key idea is to allocate the limited measurement budget to fewer unique circuits, each corresponding to a group of commuting Pauli observables.  Specifically, when the total number of measurements is limited, minimizing the confidence bound tends to allocate more resources to the high-weight groups, i.e., groups containing a larger number of commuting Pauli observables or with large coefficients, while potentially omitting measurements on some low-weight groups. This selective measurement strategy effectively minimizes the variance in the energy estimation. Notably, ROGS can significantly reduce the number of circuits prepared for measurements, which is a critical cost factor for tasks running on modern quantum computers.

The resource allocation strategy in ROGS can be interpreted using the concept of mean-field theory. In this context, the high-weight groups can be viewed as capturing the dominant mean-field interactions within the quantum system, while the low-weight groups represent local fluctuations that deviate from the mean-field approximation. By allocating more measurement resources to the larger mean-field subspaces (high-weight groups), ROGS prioritizes the measurement of the most significant contributions to the overall Hamiltonian. In contrast, distributing the measurements evenly across all groups, including those with fewer observables (low-weight groups), may lead to a less accurate representation of the system, as these groups may contain less significant interactions or local fluctuations. ROGS could have significant implications for VQE. In VQE, the objective function is the energy of a target system, which is the expectation value of the system's Hamiltonian expressed as a weighted sum of Pauli operators. The gradient calculation for the variational quantum circuit in VQE can be completed using the parameter-shift approach \cite{wierichs2022general}, which involves estimating the energy for the circuit under different parameter perturbations. Evidently, ROGS can be applied to each energy estimation procedure, thereby benefiting the gradient estimation task in VQE.

Incorporating non-QWC commuting terms \cite{yen2023deterministic} will push the boundaries of measurement reduction even further. This research direction has the potential to significantly decrease the required number of quantum measurements, making the characterization of larger-scale quantum systems more feasible. Furthermore, this approach aligns well with the mean-field \cite{izmaylov2019revising} interpretation discussed earlier. By grouping commuting terms that span across different qubits, we can capture longer-range correlations and interactions within the system. This may lead to a more accurate representation of the system's properties while still benefiting from the measurement efficiency gains provided by the overlapped grouping protocol.

Although we refer to our method as a shadow method, which acts as a pre-processing stage of classical shadow protocol to produce deterministic measurement bases before the quantum measurements, the techniques introduced here have potential applications beyond classical shadows. The optimal allocation of measurement resources and the Max-Min grouping strategy could be beneficial in other areas of quantum computing, such as VQE and gradient estimation, especially considering that ROGS could be used in energy estimation for any state including the ground state. These techniques may help enhance the efficiency of algorithms involving weighted sums of Pauli observables, similar to the Hamiltonian in the shadow method. Therefore, the insights gained from this work may have broader implications for various quantum computing applications.

While our protocol substantially reduces the number of measurement circuits and hence some classical overhead, it does not inherently account for hardware noise that can arise in NISQ devices. In particular, noise and gate imperfections can distort measurement outcomes in the low-shot regime, which is precisely the scenario where adaptive shot-allocation methods (see~\Cref{sec:fine grained shots allocation}) are most sensitive. Consequently, one may need to incorporate additional error mitigation tools — such as zero-noise extrapolation or probabilistic error cancellation — to preserve the fidelity of the allocated measurements. These methods can be integrated into our grouping-and-allocation framework so that the benefits of reduced circuit count and targeted measurement remain effective despite device-level imperfections.

\section*{Acknowledgments}

The authors thank Katharine Hyatt, Eric Kessler, Péter Kómár, Cedric Yen-Yu Lin, and Jiayu Shen for inspiring discussions and valuable input. Special thanks to Eric Kessler for his meticulous review, valuable advice, and numerous revisions.

\bibliographystyle{quantum}
\bibliography{references}

\appendix
\onecolumn

\vspace{1.2in}
\section{Confidence bound}\label{subsec:various conf-bounds}
\subsection{Proof of Theorem \ref{theorem:Confidence bound}}
In this section, we will provide a comprehensive proof of Theorem \ref{theorem:Confidence bound}, which establishes the confidence bound for accurately predicting the expectation value of the Hamiltonian. 

\begin{proof}

Based on triangle inequality, one has the Hamiltonian upper bounded by 
\begin{align}
    \left| \hat{{\mathcal{H}}} -\langle{\mathcal{H}}\rangle_{\rho}\right|
    =\left|\sum_{\ell} a_\ell(\hat{O}_\ell -\langle O_\ell\rangle_{\rho})\right|
    \leq \sum_{\ell} |a_\ell|\left|\hat{O}_\ell -\langle O_\ell\rangle_{\rho}\right|
    \leq (\sum_{\ell} |a_\ell|)\max_{\ell}\left|\hat{O}_\ell -\langle O_\ell\rangle_{\rho}\right|
\end{align}

   It is equal to prove that the maximum group's expectation value is upper bounded, i.e.
    \begin{align}\label{confidence bound 1}
&\mathrm{Pr}\left[\max_\ell\left|\hat{O}_\ell -\langle O_\ell\rangle_{\rho}\right|\geq\epsilon/\sum_\ell|a_\ell|\right]
=\mathrm{Pr}\left[\bigcup_\ell\left|\hat{O}_\ell -\langle O_\ell\rangle_{\rho}\right|\geq\epsilon/\sum_\ell|a_\ell|\right]\nonumber\\
&~~~~~~~~~~~~~~~~~~~~~~~~~~~~~~~~~~~~~~~~~~~~~~~~~~~~~~~~~~~
\leq \sum^{L}_{\ell=1}\mathrm{Pr}\left[\left| \hat{O}_\ell -\langle O_\ell\rangle_{\rho}\right|\geq\epsilon/\sum_\ell|a_\ell|\right]
    \end{align}
    The inequality arises from the union bound – also known as Boole’s inequality – which states that the probability associated with a union
of events is upper bounded by the sum of individual event probabilities.
      We can evaluate the individual group error rate probability separately. We evaluate the expectation of operators $\{O_\ell\}$ by averaging over the measurement results of the circuit. Let ${{q_j[m]}_{j\leq n}}_{m\leq M_\alpha}$ is a list of computational basis measurement results on each qubit, where $j$ is the qubit site and $m$ is the $m$th measurement, then the shadow estimation expectation value of operator $O_\ell$ is defined by average overall measurement results on this operator: $\hat{O}_\ell =M_\alpha^{-1}\sum_{m=1}^{M_\alpha} s_m^{(\ell)}$, where $s_m^{(\ell)}$ is a sign defined by the z-basis measurement result on each site within the support of that observable ${\pm 1}\ni s_m^{(\ell)}:=\prod_{j\in\mathrm{supp}\{O_\ell\}}q_j[m]$. One could expect that we can get the operator exact expectation value by taking infinite measurement: $\langle O_\ell\rangle_\rho=\lim_{M_\alpha\rightarrow\infty}\hat{O}_\ell\equiv\mathbb{E}(s_m^{(\ell)})$, then the individual term of the sum in (\ref{confidence bound 1}) can be written as: 
    \begin{align}
%\mathrm{Pr}\left[\left|\hat{O}_\ell -\langle O_\ell\rangle_{\rho}\right|\geq\epsilon/\sum_\ell|a_\ell|\right] = 
\mathrm{Pr}\left[\left|\frac{1}{M_\alpha}\sum_{m=1}^{M_\alpha} (s_m^{(\ell)}-\mathrm{E}[s_m^{(\ell)}])\right|\geq\frac{\epsilon}{\sum_\ell|a_\ell|} \right]  \leq 
2\sum_\ell \exp\left(-\frac{\epsilon'^2 M_\ell}{2}\right)
    \end{align}
    The last inequality follows from Hoeffding's inequality since each measurement result can be taken as an independent event and independent random variables, $ s_m^{(\ell)}\in\{-1,1\}$. The number of measurements performed on each operator is the sum of the shots allocated to groups to which the operator belongs, i.e., $M_\ell = \sum_{\alpha, O_\ell \in \mathcal{C}^{\mathrm{MM}}_\alpha} M_\alpha$; adopting the indicator $\mathrm{idx}_{\ell,\alpha}$ as defined in Theorem~\ref{theorem:Confidence bound}, we have $M_\ell = \sum_{\alpha=1}^A \mathrm{idx}_{\ell,\alpha} M_\alpha =M\sum_{\alpha=1}^A \mathrm{idx}_{\ell,\alpha} w_\alpha  $, where $w_\alpha = M_\alpha/M$ is the shots weight of group $\mathcal{C}_\alpha^{\mathrm{MM}}$.

  Therefore, one can accurately predict Hamiltonian:  
  \begin{align}
    \left| \hat{{\mathcal{H}}} -\langle{\mathcal{H}}\rangle_{\rho}\right|\leq \epsilon
\end{align}
with accuracy $1-\delta$, where
\begin{align}\label{eqn.conf-bound}
    \delta = 2\sum_{\ell=1}^{L} \exp\left(-\frac{\epsilon^2 M \sum_\alpha \mathrm{idx}_{\ell,\alpha}w_\alpha}{2(\sum_\ell|a_\ell|)^2}\right)\geq\mathrm{Pr}\left[\max_{\ell}\left|\hat{O}_\ell -\langle O_\ell\rangle_{\rho}\right|\geq\epsilon/(\sum_\ell |a_\ell|)\right],~~\sum_\alpha w_\alpha =1.
\end{align}
    
\end{proof}
\vspace{-0.1in}
\textbf{Remarks.}~We can find a shot allocation $\{M_\alpha\}$ is allocated among the overlapped groups $\{\mathcal{C}_\alpha^{\mathrm{MM}}\}$ given the total number of measurement budgets $M$, where $M = \sum_\alpha M_\alpha$, to minimize the confidence bound~\ref{theorem:Confidence bound} through convex optimization of a Log-Sum-Exp type function. However, note that although $\epsilon$ is the error bound in the tail bounds, it functions as a hyperparameter in the convex optimization problem, meaning that for each $\epsilon$, there is corresponding minimum confidence bound $\delta$ as indicated in (\ref{eqn.conf-bound}), while the optimized result, or shots allocation $\{w_\alpha\}$, remains unchanged provided that $\epsilon^2 M$ is constant. Therefore, one should aim to find the optimal $\epsilon_0$ for a small budget $M_0$, keeping the hyperparameter $\epsilon_0^2 M_0$ fixed, i.e., for any total budget $M$, one would choose $\epsilon = \epsilon_0\sqrt{M_0/M}$ to achieve the optimal measurement result, which is the core concept of algorithm~\ref{alg:grouping-rogs-fixed}. 

Eq.~\eqref{eqn.conf-bound} gives confidence bound for the estimation of the Hamiltonian's expectation value when each observable $O_{l}$ is estimated with $M_\ell =M\sum_{\alpha=1}^A \mathrm{idx}_{\ell,\alpha} w_\alpha  $ measurement shots. On the other hand, as the total number of shots $M$ is fixed, the various choices of the weights $\{w_{\alpha}\}$ lead to different $\delta$. As the lower $\delta$ leads to higher confidence, we want to find the optimal choice of $\{w_{\alpha}\}$ under constraint $\sum w_{\alpha} = 1$ to minimize the hyperparameter $\delta$.

\subsection{Comparison between confidence bound}\label{subsec:group Hoeffding Bound}
This subsection introduces a supplementary confidence bound. This modification from the previously discussed confidence bound aims to address scenarios where groups with larger numbers of terms but smaller average coefficients are compared against groups with fewer terms but larger coefficients. 

Apparently, the Hamiltonian can be grouped by $\mathcal{C}^{\mathrm{MM}}_\alpha$
\begin{align}
      \left| \hat{{\mathcal{H}}} -\langle{\mathcal{H}}\rangle_{\rho}\right|=\left|\sum_{\alpha =1}^A\sum_{O_\ell\in\mathcal{C}^{\mathrm{MM}}_\alpha} a_\ell(\hat{O}_\ell -\langle O_\ell\rangle_{\rho})\right|
      \leq \sum_{\alpha =1}^A\left|\sum_{O_\ell\in\mathcal{C}^{\mathrm{MM}}_\alpha} a_\ell w^o_{\ell,\alpha}(\hat{O}_\ell -\langle O_\ell\rangle_{\rho})\right|
\end{align}
where $w^o_{\ell,\alpha}$ is a weight distribution such that $w^o_{\ell,\alpha}\geq 0$, $\sum_{\alpha}w^o_{\ell,\alpha} = 1$. One example is $w^o_{\ell,\alpha}= (\sum_{\alpha}\mathrm{idx}_{\ell,\alpha})^{-1}$. Therefore, one could apply a tail bound to the max-min group instead of the individual operators of the Hamiltonian. Similar to the previous section, one can upper-bound the error of the Hamiltonian by demonstrating that the maximum group's expectation value has a high confidence level:
    \begin{align}\label{confidence bound --group level}
\mathrm{Pr}\left[\max_\alpha\left|\sum_{O_\ell\in\mathcal{C}^{\mathrm{MM}}_\alpha}\!\! {a_\ell}{w^o_{\ell,\alpha}}(\hat{O}_\ell -\langle O_\ell\rangle_{\rho})\right|\geq\epsilon/A\right]
\leq \sum^{A}_{\alpha=1}\mathrm{Pr}\left[\left|\sum_{O_\ell\in\mathcal{C}^{\mathrm{MM}}_\alpha}\!\! a_\ell{w^o_{\ell,\alpha}}(\hat{O}_\ell -\langle O_\ell\rangle_{\rho})\right|\geq\epsilon/A\right]
    \end{align}
    based on the same argument as the previous section, one has
       \begin{align}
&\mathrm{Pr}\left[\left|\sum_{O_\ell\in\mathcal{C}^{\mathrm{MM}}_\alpha}\frac{{a_\ell}{w^o_{\ell,\alpha}}}{M_\alpha}\sum_{m=1}^{M_\alpha} (s_m^{(\ell)}-\mathrm{E}[s_m^{(\ell)}])\right|\geq\frac{\epsilon}{A} \right] \nonumber\\&=\mathrm{Pr}\left[\frac{1}{M_\alpha}\left|\sum_{m=1}^{M_\alpha} \left(\sum_{O_\ell\in\mathcal{C}^{\mathrm{MM}}_\alpha}{a_\ell}{w^o_{\ell,\alpha}} s_m^{(\ell)}-\mathbb{E}\left[\sum_{O_\ell\in\mathcal{C}^{\mathrm{MM}}_\alpha}{a_\ell}{w^o_{\ell,\alpha}} s_m^{(\ell)}\right]\right)\right|\geq\frac{\epsilon}{A} \right] 
    \end{align}
    considered that each measurement result can be taken as an independent event and independent random variables $(X_m)_{m\leq M_\alpha}$ are bounded s.t. $X_m :=\sum_{O_\ell\in\mathcal{C}^{\mathrm{MM}}_\alpha}{a_\ell}{w^o_{\ell,\alpha}} s_m^{(\ell)}\in[-\sum_{O_\ell\in\mathcal{C}^{\mathrm{MM}}_\alpha}{|a_\ell|}{w^o_{\ell,\alpha}},\sum_{O_\ell\in\mathcal{C}^{\mathrm{MM}}_\alpha}{|a_\ell|}{w^o_{\ell,\alpha}}$, therefore based on Hoeffding's inequality, each group error probability is bounded by:
    \begin{align}\label{eqn.group Hoeffding inequality}
        \mathrm{Pr}\left[\frac{1}{M_\alpha}\left|\sum_{m=1}^{M_\alpha} \left(X_m-\mathbb{E}\left[X_m\right]\right)\right|\geq\frac{\epsilon}{A}  \right] \leq 2\exp\left(-\frac{\epsilon^2 M_\alpha}{2(A\sum_{O_\ell\in \mathcal{C}^{\mathrm{MM}}_\alpha
        }{|a_\ell|}{w^o_{\ell,\alpha}})^2} \right)
    \end{align}
    by taking summation over all groups as in the previous section, one could obtain the confidence bound~(\ref{eqn.conf-bound}). This revised bound could provide a more equitable distribution among groups compared to the initial one, depending on the setting of weight parameter $w^o_{\ell,\alpha}$. If the measurement budget is not a constraint, this form of the confidence bound could be instrumental in deciding how shots are allocated among groups. For instance, by setting $w^o_{\ell,\alpha}$ as a constant concerning $\alpha$, i.e. $w^o_{\ell,\alpha}=1/\sum_\alpha\mathrm{idx}_{\ell,\alpha}$. In general, one might find beneficial results by setting it proportional to the number of operators within the respective group $w^o_{\ell,\alpha}\propto \sum_\ell \mathrm{idx}_{\ell,\alpha}$ or even setting $w^o_{\ell,\alpha} = 0$ for all low-weighted groups (groups with a small number of operators or small absolute values of coefficients). This approach allows the operators to concentrate on high-weight groups, thereby expanding the main-field subspace. With finite quantum resources, this strategy helps focus resources on the most significant interactions within the system, thus dealing effectively with limited resources (see \Cref{subsec:shots allocation}). 

   ~\\
\textbf{Remarks.}~Interestingly, the hyper-parameter $\epsilon^2 M$ can be interpreted as an inverse temperature in a thermal state. At high temperatures, analogous to thermal agitation, the shots distribute more uniformly across the groups, resembling a system with a flattened Boltzmann distribution. Conversely, at low temperatures, the system approaches its thermal ground state, and the shots predominantly concentrate in high-weight groups, mimicking the population of low-energy states. This behavior is reminiscent of the Boltzmann distribution, where the probability of occupying a state decreases exponentially with its energy. By tuning the hyper-parameter $\epsilon^2 M$, one can effectively control the "temperature" of the shot allocation, allowing for a balance between exploration and exploitation of the group weights.
    
    Although this version of the confidence bound takes the effect of coefficients on operators into account, its overemphasis on coefficients can lead to less accurate estimations. This inaccuracy may arise from factors beyond just the coefficients; for example, the variance of the operators, which is state-dependent, also affects the estimation accuracy. An extreme example is the identity operator in a qubit Hamiltonian, which always has the largest coefficients among all operators but zero variance. Therefore, one can obtain the expectation value of the identity operator without any measurement, which is why we consider only the traceless Hamiltonian. The same argument applies to all operators: the variance does not depend on the coefficients but is instead state-dependent. Therefore, simply considering the coefficients as the measurement weight does not yield satisfactory results. It could be beneficial to take the variance of operators into account when we have some knowledge of the state after initial measurements (see \Cref{subsec:adaptive grouping shadow}).

    The supplementary confidence bound involves a more complex optimization process, where the weights $w^o_{\ell,\alpha}$ are adjusted according to the operator coefficients and the structure of the Hamiltonian. While this allows for a tailored approach to measurement allocation, it also introduces a higher risk of overfitting to specific dataset characteristics. This complexity can make the bound less robust, especially in scenarios where the underlying quantum state or the Hamiltonian's properties are not perfectly understood or are subject to variations. Although this confidence bound might appear effective, it does not consistently yield superior results across different scenarios. The simpler bounds in the main paper, though less tailored, provide more stable and predictable performance across a wider range of conditions, making them more reliable for general use. 
\subsubsection{A Toy Model}\label{subsection:conter exp}
Consider an $n$-qubit system with a Hamiltonian of the form:
\begin{align}
    {\mathcal{H}} &=  \bigotimes_{i=1}^n\sigma_i^z ~+ ~2^{-2n}\sum_{P \in \{I, x\}} \bigotimes_{i=1}^n\sigma_i^P
\end{align}
where $\sigma_i^x, \sigma_i^y, \sigma_i^z, \sigma_i^I$ are Pauli operators acting on the $i$-th qubit of the system. In this Hamiltonian, all operators with coefficients $2^{-2n}$ are qubit-wise commuting (QWC) with each other. The operators in ${\mathcal{H}}$ form two qubit-wise commuting groups, which can be denoted by
\begin{align}
    \mathcal{C}_1^{\text{MM}} = \left\{\bigotimes_{i=1}^n\sigma_i^z,~ \mathbb{I}_{2^n}\right\}, \qquad
    \mathcal{C}_2^{\text{MM}} = \bigcup_{P \in \{I, x\}} \bigotimes_{i=1}^n\sigma_i^P
\end{align}
with measurement gates 
\begin{align}
     \mathcal{P}_1 = \bigotimes_{i=1}^n\sigma_i^z, \qquad
     \mathcal{P}_2 = \bigotimes_{i=1}^n\sigma_i^x
\end{align}
respectively for each group. The size of group 1 is fixed at $|\mathcal{C}_1^{\text{MM}}| = 2$, and the size of group 2 increases exponentially with the size of the system $n$, $|\mathcal{C}_2^{\text{MM}}| = 2^n$. Evidently, the size of $\mathcal{C}_2^{\text{MM}}$ is much larger than that of $\mathcal{C}_1^{\text{MM}}$ as $n$ increases, i.e., $|\mathcal{C}_2^{\text{MM}}| \gg |\mathcal{C}_1^{\text{MM}}|$. However, the sum of the absolute values of the coefficients\footnote{We use the sum of the absolute values of the coefficients $\sum_{O_\ell \in \mathcal{C}^{\text{MM}}}|a_\ell|$ as an indicator based on \Cref{eqn.group Hoeffding inequality}.} associated with operators in $\mathcal{C}_1$ (order of $O(1)$) is significantly greater than that of $\mathcal{C}_2$ (order of $2^{-n}$).

In this scenario, the confidence bound provided in Theorem~\ref{theorem:Confidence bound} may not provide the most optimal allocation of measurement resources. This is because it only takes into account the sizes of the groups but not the magnitudes of the coefficients associated with the operators within each group. Consequently, Theorem~\ref{theorem:Confidence bound} would suggest allocating more measurement resources to $\mathcal{C}_2$ due to its larger size. However, this allocation may not be optimal because the operator in $\mathcal{C}_1$, despite being a single term, has a much higher coefficient and thus contributes more significantly to the overall Hamiltonian.
In such cases, the supplementary confidence bound introduced in Subsection~\ref{subsec:group Hoeffding Bound} can be more useful. This alternative bound, \eqref{eqn.group Hoeffding inequality}, takes into account the coefficients of the observables within each group, allowing for a more nuanced allocation of measurement resources. By considering each group's relative importance based on its size and the magnitudes of its observable coefficients, the supplementary confidence bound can help mitigate the potential breakdown of the primary bound in these extreme scenarios.

Notably, while Theorem~\ref{theorem:Confidence bound} may not provide the most optimal allocation of measurement resources without sampling the hyperparameter space, the breakdown of this theorem in this scenario does not imply that the theorem is invalid or unusable. Instead, it suggests that a more nuanced approach to determining the hyperparameter $\epsilon$ is necessary. As discussed earlier, $\epsilon$ influences the distribution of measurement shots among the groups. As $\epsilon$ decreases, the optimal shot allocation $w_\alpha$ tends to concentrate on the high-weight groups; conversely, as $\epsilon$ increases, the shot distribution becomes more evenly spread across all groups. This example indicates that when dealing with a Hamiltonian, if such an imbalance between size and coefficients occurs in some extreme cases, one would be better off starting with a large $\epsilon$.

\subsection{Adaptive Resource-Optimized Grouping Shadow}\label{subsec:adaptive grouping shadow}
This subsection introduces an adaptive version of the ROGS method, which incorporates prior information about the quantum state to refine the measurement resource allocation strategy. Given that our knowledge of the circuits accumulates in the process of measurement, such as the entanglement between the qubits in the circuits will affect the variance of the expectation value of operators. This allows for a sharper confidence bound by shifting from Hoeffding (we used in the proof of Theorem \ref{theorem:Confidence bound}) to Bernstein inequality in our calculations,

\begin{theorem}
Given a set of overlapping groups $\{\mathcal{C}^{\mathrm{MM}}_\alpha\}_\alpha$ for a collection of operators $\{O_\ell\}$ from the Hamiltonian ${\mathcal{H}} = \sum_\ell a_\ell O_\ell$, and a measurement weight allocation for each group $\{w_\alpha\}$, we obtain statistical results for the operators, namely variances $(\mathrm{Var}[\hat{O}_\ell])_\ell$. Then, the confidence bound is defined as
\begin{align}\label{conf-bound_Bernstein}
    &\mathrm{Conf}\left[\epsilon,\{w_\alpha\};M,\{\mathcal{C}_\alpha^{\mathrm{MM}}\}\right] := 2\sum_{\ell=1}^{L} \exp\left(-\frac{\epsilon^2 M\sum_{\alpha}\mathrm{idx}_{\ell,\alpha}\;w_\alpha }{2(\sum_\ell|a_\ell|)^2(\mathrm{Var}[\hat{O}_\ell]+\epsilon\max_\ell(\hat{O}_\ell)/3)} \right) = \delta
\end{align}
with constrain condition $\sum_\alpha w_\alpha = 1$, $w_\alpha\geq 0$ for all $\alpha$; where the indicator $\mathrm{idx}_{\ell,\alpha}$ is defined as
\begin{align}
\mathrm{idx}_{\ell,\alpha}=\left\{
        \begin{array}{ll}
        1 & \text{if } O_\ell \in \mathcal{C}^{\mathrm{MM}}_\alpha\nonumber\\
         0 & \text{otherwise} 
        \end{array}\right.
\end{align}

Then, a collection of $M$ independent classical shadows enables accurate prediction of the expectation of the Hamiltonian for a given quantum state $\rho$:
\begin{align}
        \left| \hat{{\mathcal{H}}} -\langle{\mathcal{H}}\rangle_{\rho}\right|\leq \epsilon,
\end{align}
with probability $1-\delta$.

\end{theorem}

The theorem illustrates the robust efficacy of the proposed tighter confidence bound, especially when prior knowledge about the operators for specific states is incorporated. Consequently, we propose the subsequent algorithm for state-adaptive ROGS:

 \begin{algorithm}[H]
{\small
\begin{algorithmic}[1]
\caption{{\small \textbf{(State-Adaptive ROGS)} %(high-level)
}}
\label{alg:adaptive grouping shadow}

\Require
Measurement budget $M$ and Hamiltonian ${\mathcal{H}} = \sum_{\ell}a_{\ell}O_\ell$. 

\Ensure
A fixed Pauli basis measurement recipe $\{\mathcal{P}_\alpha^{\times M_\alpha}\}$ containing $\{X,Y,Z\}^{n\times M}$ for the classical shadow quantum measurement stage. 

\vspace{.1in}

\State Identify the max-min groups $\{C_\alpha^{\mathrm{MM}}\}$ of operators $\{O_\ell\}$ in ${\mathcal{H}}$ using Algorithm~\ref{alg:MM-grouping}.
\vspace{.1in}
\For{$t=1$ to $T$}\Comment{$M = \sum_t M^t$}

\State Pre-processing: \begin{itemize}
    \item[\textit{1}]. $\{\omega_{\alpha}^{t}\}^* = \arg\min_{\{w_\alpha\}} \mathrm{Conf}\left[\epsilon,\{w_\alpha\},\{\mathrm{Var}_{\sum_{i=1}^{t-1}M_\alpha^{i}}[\mathcal{C}^{\mathrm{MM}}_\alpha]\}_\alpha,M^t,{\mathcal{H}}\right]$
    \item[\textit{2}]. $\delta^{t*} = \arg\min_{\delta} \mathrm{Error}(M^t_{\mathrm{ini}},\{\omega^t_\alpha\}^*(\delta; {\mathcal{H}}))$
    \item[\textit{3}]. Get recipes$_t\ni \{X,Y,Z\}^{n\times M^t}$
    \end{itemize}
\State Quantum Measurement: perform $z-$basis measurement based on recipes$_t$
\State Post-processing: evaluate eigenvalue $\langle {\mathcal{H}}\rangle_{M^t}$, and variance for all $t$-measurements $(\mathrm{Var}_{\sum_{i=1}^t M_\alpha^{i}}[\mathcal{C}^{\mathrm{MM}}_\alpha])_\alpha$

\EndFor
\end{algorithmic}
}
\end{algorithm}
This approach will necessitate a hybrid interplay between classical (steps 2 and 4) and quantum (step 3) processes, with an imperative to curtail the number of circuits utilized. Addressing this challenge will be the focus of our future work. The adaptive approach is particularly relevant for scenarios where multiple measurements are performed on closely related quantum states, such as in the VQE algorithm.

\section{Median-of-means estimation}\label{subsec:MoM}
This section discusses the median-of-means estimator and its advantages over the simple mean estimator for heavy-tailed distributions. After allocating measurement shots to max-min groups by minimizing the confidence bound (\ref{conf-bound}), one can achieve high hit rates for high-weight operators (see Fig.\ref{fig:GSE error} right). This is considering that all Pauli operators $O_\ell = \prod_{i\in\mathrm{supp}(O_\ell)} P_i$ can yield only two measurement outcomes $(-1,1) \ni s_{m}^{(\ell)}:=\prod_{i\in\mathrm{supp}(O_\ell)} q_i[m]$ (where $m$ denotes the $m$-th measurement, and $q_i[m]$ represents the z-basis measurement result at the $i$-th site). Consequently, all measurement results for an operator are i.i.d samples and locally conform to a sub-Gaussian (binomial) distribution. It has been demonstrated that the median of the mean estimator
\begin{align}\label{eqn.mom estimator}
    \hat{O}_\ell(K_\ell) = \mathrm{median}[\hat{O}^{1}_\ell,\cdots, \hat{O}^{K_\ell}_\ell],\qquad \hat{O}^I_\ell := \frac{1}{N_\ell}\sum_{m = I N_\ell +1}^{(I+1) N_\ell} s_m^{(\ell)};~~ N_\ell = \lfloor M_\ell/ K_\ell\rfloor
\end{align}
will yield a more accurate overall mean estimation $\langle O_\ell\rangle_\rho$ than the mean estimator $\hat{O}_\ell = \frac{1}{M_\ell}\sum_{m=1}^{M_\ell}s_m^{(\ell)}$. 

\begin{facts}
Given an i.i.d. sample $(X_1,\dots,X_M)$ from a distribution on $\mathbb{R}$ with mean $\mu$ and finite variance $\sigma^2 < \infty$, 
for any $K,\epsilon > 0$, 
\begin{align}
   \Pr\left( \left|\mu_{\mathrm{MoM}}(K) - \mu\right| > \epsilon\right) \leq \exp\left(-\frac{K}{2}\left(1 - 2\frac{K\sigma^2}{M\epsilon^2}\right)^2\right)
\end{align}
where $\mu_{\mathrm{MoM}}(K)$ is the median-of-means estimator and $K \leq M$ is the number of groups into which the samples are evenly split.
\end{facts}

The choice of the number of groups $K$ in the median-of-means estimator involves a bias-variance trade-off. As $K$ increases, the bias of the estimator decreases, but the variance increases. Conversely, as $K$ decreases, the bias increases, but the variance decreases. The optimal choice of $K$ balances this trade-off and depends on the sample size $M$. If $M$ is large, a larger number of groups can be used without compromising the accuracy of the estimator. In practice, a common choice for $K$ is to set it proportional to the square root of the sample size, i.e., $K = \sqrt{M}$. This choice often provides a good balance between bias and variance, especially when the sample size is sufficiently large. However, it's important to note that the specific choice of $K$ may also depend on the characteristics of the underlying distribution and the desired level of accuracy. In the numerical results, for each Pauli operator $O_\ell$, we have adopted $K_\ell = M_\ell \sqrt{\sigma_\ell^2/\epsilon_\ell^2}$, where $\epsilon_\ell$ is related to $\epsilon$ in the confidence bound~\ref{theorem:Confidence bound} by $\epsilon_\ell = \epsilon/\sum_\ell |a_\ell|$. The $\sigma_\ell$ has been obtained by the variance of the operator $O_\ell$, which has been determined from $M_\ell$ quantum measurements $\hat{O}_\ell$. Here, $M_\ell$ is the number of measurements performed on $O_\ell$, which was decided by the shots allocation $M_\ell = M\sum_{\alpha} \mathrm{idx}_{\ell,\alpha}w_\alpha$. 

The median-of-means estimator serves as a superior mean estimator in our context due to the shots allocation determined by our Algorithm~\ref{alg:grouping-rogs-fixed}, which concentrates measurements on the high-weight groups. Consequently, when the measurement budget is limited, our selective measurement approach focuses on the high-weight operators rather than distributing measurements evenly among all operators. This targeted allocation ensures that each operator receives a sufficient number of shots, enabling the median-of-means estimator to perform more effectively. To illustrate the benefits of the median-of-means estimator in conjunction with our shots allocation strategy, we compare the results for \chem{N H_3} (JW encoding) with and without the median-of-means estimation under a limited total number of measurements.

\begin{figure}[htbp]
    \centering
\includegraphics[width=0.35\textwidth]{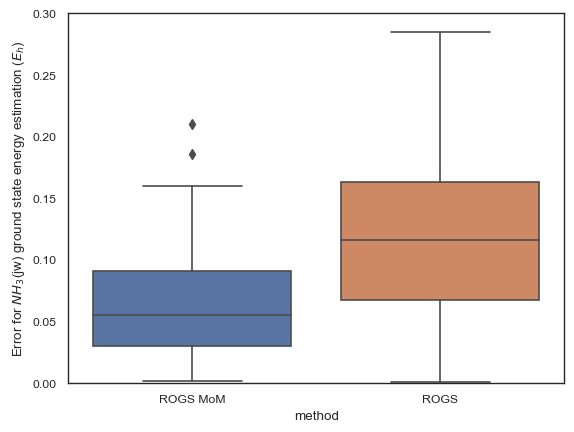}
    \caption{\small{\textit{Advantage of the Median-of-means estimation for \chem{NH_3} Using Jordan-Wigner~\cite{jordan1993paulische} (JW) Encoding:} This plot compares the ground state estimation results for the same molecule under the same settings when limited to 1000 total measurements, except that \textit{ROGS MoM} employs the median-of-means estimator, highlighting its benefits. Here, we have set $K = M_\ell{\epsilon_\ell^2}/{\sigma_\ell^2}$, where $\epsilon_\ell = \epsilon/\sum_\ell |a_\ell|$, $\epsilon$ is the error bound as given by the confidence bound~\ref{theorem:Confidence bound}. }}

    \label{fig:NH_3 MoM}
\end{figure}
\vspace{-0.2in}
\section{Fine Graining Shots Allocation}\label{sec:fine grained shots allocation}
In this section, we'll discuss how to find the shots allocation in detail, specifically, how to determine the hyperparameter $\epsilon$ in (\ref{conf-bound}) when we treat it as a convex optimization problem, which acts as the error bound of energy estimation of a particular state in (\ref{eqn:The:error bound of energy estimation}). As observed in the confidence bound expression~(\ref{conf-bound}) in Theorem~\ref{theorem:Confidence bound}, the shot allocation, resulting from the minimization of the confidence bound, remains the same provided the product of the state energy estimation error bound and the total budget, $\epsilon^2 M$, is kept constant. Assuming $\epsilon \leq \epsilon_0 = 2\sum_{\ell}|a_\ell|$ is valid for all states, we empirically choose $\epsilon = \epsilon_0(M_0/M)^{1/2}$ for a given budget $M$, for any $M_0 \geq 0$, and determine the shot allocation by performing convex optimization on the confidence bound. This typically provides satisfactory results without necessitating the fine-tuning of the parameter $\epsilon$ for specific states, as delineated in Algorithm~\ref{alg:grouping-rogs-fixed}. By setting $M_0 = 1$, we can determine the fixed-shot outcomes for our ROGS method (see error table \Cref{table:ROGS naive evenly distribution} \textit{ROGS Naive}).

\begin{table}[h!]
\scalebox{0.82}{
  \begin{tabular}{|l|ccc|ccc|ccc|ccc|ccc|}
    \hline
    \multirow{3}{*}{\makecell{Molecule\\Encoding}} & \multicolumn{3}{c|}{\makecell{\chem{H_2}\\
    (-1.86)}} & \multicolumn{3}{c|}{\makecell{\chem{LiH}\\
    (-8.91)}} & \multicolumn{3}{c|}{\makecell{\chem{BeH_2}\\(-19.04)}} & \multicolumn{3}{c|}{\makecell{\chem{H_2O}\\
    (-83.6)}} & \multicolumn{3}{c|}{\makecell{\chem{NH_3}\\(-66.88)}} \\
    \hline
      & {JW} & {P} & {BK} 
      & {JW} & {P} & {BK} 
      & {JW} & {P} & {BK} 
      & {JW} & {P} & {BK} 
      & {JW} & {P} & {BK}  \\
    \hline
        ROGS Naive & 
        0.04 & 0.02 & 0.04 &
        0.03 & 0.02 & 0.01 & 
        0.06 & 0.08 & 0.05 & 
        0.10 & 0.09 & 0.13 & 
        0.15 & 0.13 & 0.11\\
    \hline
     Even distribution&
        0.07 & 0.05 & 0.05 &
        0.10 & 0.05 & 0.06 &
        0.10 & 0.11 & 0.12 &
        0.16 & 0.22 & 0.31 &
        0.31 & 0.38 & 0.32\\
    \hline
  \end{tabular}}
  \caption{\small{\textit{Naive Estimation benchmark:} This table shows the RMSE ($E_h$) of Naive shots distribution given by a fixed empirical error bound $\epsilon$ without pre-sampling \textit{(ROGS Naive)}, and the RMSE given by distributing the measurements evenly across all groups \textit{(Even distribution)} by simulating the process 10 times, each with 1,000 measurements.}}
\label{table:ROGS naive evenly distribution}
\vspace{-.6em}
\end{table}

These results underscore the method's robustness within a certain range of the parameter $\epsilon$. On the other hand, if one holds the product $\epsilon^2 M$ fixed, the shots allocation ${w_\alpha}$ obtained by the convex optimization is fixed, and so is the confidence bound $\delta = \mathrm{Conf}\left[\epsilon_0,{w_\alpha};M_0,{\mathcal{H}}\right]$, which implies that this hyperparameter $\epsilon_0$ can be ascertained from a small number of measurements $M_0$. With $\epsilon_0^2 M_0$ held fixed in ~(\ref{conf-bound}), the shot allocation $\{w_\alpha\}$ which minimizes the confidence bound is unchanged, meaning that if we increase the total number of shots to $M$ while setting $\epsilon = \epsilon_0\sqrt{M_0/M}$, the shots allocation $\{w^\alpha\}$ which minimizes the confidence bound $\delta = \mathrm{Conf}\left[\epsilon_0,{w^*_\alpha};M_0,{\mathcal{H}}\right]$ remains unchanged. Therefore, the algorithm can be improved by incorporating a coarse-grained search for the hyperparameter $\epsilon_0$ to find the minimum error of the energy estimation for a given state with a finite number of shot measurements $M_0$, where the energy estimation $\hat{\mathcal{H}}$ is obtained by quantum measurement using the measurement recipe $\{\mathcal{P}_\alpha^{\otimes M_{\alpha}}\}_{1\leq \alpha\leq A}\in\{X,Y,Z\}^{n\times M_0},~ (\sum M_\alpha^0 = M)$, which is given by minimizing the confidence bound using a given $\epsilon_0$ following algorithm~\ref{alg:grouping-rogs-fixed},
\begin{align}\label{eqn.state error estimation}
    \mathrm{Error}(\{\mathcal{P}_\alpha^{\otimes M_alp}\})= \sqrt{\mathbb{E}\left[\left(\hat{{\mathcal{H}}}(\{\mathcal{P}_\alpha^{\otimes M_\alpha}\}_{1\leq \alpha\leq A}) -\langle{\mathcal{H}}\rangle_{\rho}\right)^2\right]};
\end{align}
then applying $\epsilon = \epsilon_0(M_0/M)^{1/2}$ for the total amount of shots $M$. Here is our detailed algorithm for determining the optimal shot allocation:

\begin{algorithm}[H]
{\small
\begin{algorithmic}[1]
\caption{{\small \textbf{(Coarse-Graining Parameter ROGS)}}}
\label{alg:coarse-graining ROGS}
\Require
Measurement budget $M$ and Hamiltonian ${\mathcal{H}} = \sum_{\ell}a_{\ell}O_\ell$. 

\Ensure
A fixed Pauli basis measurement recipe $\{\mathcal{P}_\alpha^{\times M_\alpha}\}$ containing $\{X,Y,Z\}^{n\times M}$ for the classical shadow quantum measurement stage. 

\vspace{.1in}

\State Identify the max-min groups $\{C_\alpha^{\mathrm{MM}}\}$ of operators $\{O_\ell\}$ in ${\mathcal{H}}$ using Algorithm~\ref{alg:MM-grouping}.
\State initialize $error = {}$
\While{$i\leq N$} \Comment{Coarse-Graining with $M_{test}$ shots. Use $M_{cg}=M_{test}/N$ shots in each rounds}
    \State $M_0 = random(0.1, 10)$
    \State $\epsilon_{cg} = 2\sum_\ell|a_\ell|\sqrt{M_0/M_{cg}}$, total shots $=~M_{cg}$, run algorithm~\ref{alg:grouping-rogs-fixed}. \textbf{Return} recipe $\{\{\mathcal{P}^{(m)}_\alpha\}_{1\leq m\leq M_\alpha}\}$ 
    \State $error[\epsilon_{cg}]=\mathrm{Error}(\{\{\mathcal{P}^{(m)}_\alpha\}_{1\leq m\leq M^{cg}_\alpha}\})$ (\ref{eqn.state error estimation})
\EndWhile
\State  $\epsilon_{cg}^* = \min(error[\epsilon_{cg}]).key()$\Comment{get optimal error rate for shots $M_{cg}$}
\State $\epsilon = \epsilon_{cg}\sqrt{M_{cg}/M}$  total shots $=~M$, run algorithm~\ref{alg:grouping-rogs-fixed}
\end{algorithmic}
}
\end{algorithm}
Our algorithm shows a strong advantage after going through this process. One should note the hyperparameter $\epsilon$ in \Cref{theorem:Confidence bound} influences the distribution of measurement shots among the groups. As $\epsilon$ decreases, the optimal shot allocation $w_\alpha$ tends to concentrate on the high-weight groups. Conversely, as $\epsilon$ increases, the shot distribution becomes more evenly spread across all groups. This behavior can be intuited from the role of $\epsilon$ in the confidence bound: a smaller $\epsilon$ demands higher precision, which is achieved by focusing measurements on the most significant terms in the Hamiltonian.

Interestingly, the cost function given by \Cref{subsec:group Hoeffding Bound} exhibits the opposite behavior when the weight factor $w_{\ell,\alpha}^0$ is concentrated on the high-weight groups. In this case, as the weight factors become more focused on the high-weight groups, the cost function tends to allocate shots more evenly among all groups. This is because the concentrated weight factors already prioritize the important terms, allowing the shot allocation to balance the measurement of less significant terms for overall accuracy.
These contrasting behaviors highlight the interplay between the hyperparameter $\epsilon$, the weight factors $w_{\ell,\alpha}^0$, and the resulting shot allocation. The choice of $\epsilon$ and the definition of the weight factors should be carefully considered based on the desired balance between prioritizing important terms and ensuring a comprehensive measurement of the entire Hamiltonian. Further research into the optimal selection of these parameters for specific applications could lead to more efficient and effective measurement strategies in quantum algorithms.

\end{document}